\documentclass[aps,twocolumn,superscriptaddress,nofootinbib,longbibliography,10pt,preprintnumbers]{revtex4-1}
\usepackage{graphicx}
\usepackage{amsmath}
\usepackage{amssymb}
\usepackage{bm}
\usepackage{mathtools}
\usepackage{hyperref}
\usepackage{tikz}
\usetikzlibrary{calc}
\usepackage{tikz-cd}
\hypersetup{
    linktocpage=true,
    colorlinks=true,
    citecolor=black,
    linkcolor=black,
    urlcolor=black,
}

\DeclareMathOperator{\SO}{SO}
\DeclareMathOperator{\Spin}{Spin}
\DeclareMathOperator{\U}{U}
\DeclareMathOperator{\arf}{Arf}

\DeclareMathOperator{\SU}{SU}

\DeclareMathOperator{\TY}{TY}
\DeclareMathOperator{\Rep}{Rep}
\DeclareMathOperator{\Vect}{Vect}
\DeclareMathOperator{\bfVect}{\mathbf{Vect}}
\DeclareMathOperator{\Mod}{Mod}
\DeclareMathOperator{\TQFT}{TQFT}
\DeclareMathOperator{\bfTQFT}{\mathbf{TQFT}}
\DeclareMathOperator{\Hom}{Hom}
\DeclareMathOperator{\End}{End}
\newcommand{\Sfaith}{\mathcal{S}^{\text{faith}}}

\usepackage[status=draft,inline,nomargin]{fixme}
\fxusetheme{color}
\FXRegisterAuthor{ko}{ako}{KO}

\begin{document}

\preprint{ YITP-26-16}
\preprint{ RIKEN-iTHEMS-Report-26}

\title{Symmetry Spans and Enforced Gaplessness}

\author{Takamasa Ando}
\affiliation{Center for Gravitational Physics and Quantum Information, Yukawa Institute for Theoretical Physics, Kyoto University, Kyoto 606-8502, Japan}

\author{Kantaro Ohmori}
\affiliation{RIKEN Center for Interdisciplinary Theoretical and Mathematical Sciences (iTHEMS), RIKEN, Wako, Saitama 351-0198, Japan}
\affiliation{Department of Physics, The University of Tokyo, Bunkyo-ku, Tokyo 113-0033, Japan}

\begin{abstract}
Anomaly matching for continuous symmetries has been the primary tool for establishing \textit{symmetry enforced gaplessness}---the phenomenon where global symmetry alone forces a quantum system to be gapless in the infrared.
We introduce a new mechanism based on \textit{symmetry spans}: configurations in which a global symmetry $\mathcal{E}$ is simultaneously embedded into two larger symmetries, as $\mathcal{D}\hookleftarrow\mathcal{E}\hookrightarrow\mathcal{C}$.
Any gapped phase with the full symmetry must, upon restriction to $\mathcal{E}$, arise as the restriction of both a gapped $\mathcal{C}$-symmetric phase and a gapped $\mathcal{D}$-symmetric phase.
When no such compatible phase exists, gaplessness is enforced.
This mechanism can operate with only discrete and non-anomalous continuous symmetries in the UV, both of which admit well-understood lattice realizations.
We construct explicit symmetry spans enforcing gaplessness in 1+1 dimensions, exhibit their realization in conformal field theories, and provide lattice Hamiltonians with the relevant symmetry embeddings.
\end{abstract}

\maketitle

\tableofcontents

\section{Introduction}
Determining the infrared fate of a quantum system from its ultraviolet data is a central challenge in theoretical physics.
Global symmetry provides one of the most powerful tools for this problem: 't Hooft anomaly matching \cite{tHooft:1979rat} requires that anomalies of global symmetries be preserved under renormalization group flow, thereby constraining the allowed infrared phases.
In condensed matter physics, related constraints are known as the Lieb--Schultz--Mattis theorem and its generalizations \cite{Lieb:1961fr, koma2000spectral, Oshikawa:2000zwq, roy2012space, Hastings:2003zx, Hastings_2005, Watanabe:filling, Qi2017ground, Huang2017building, Cheng:2018lti, Kobayashi:2018yuk, Else:2019lft, Ogata:2020hry, Aksoy:2021uxb, Cheng:2022sgb, Yao:2023bnj, Aksoy:2023hve, Kapustin:2024rrm, Liu:2024nox, Ma:2024sig}.
The notion of global symmetry has recently been generalized to include higher-form \cite{Kapustin:2014gua, Gaiotto:2014kfa} and non-invertible symmetries \cite{Bhardwaj:2017xup, Tachikawa:2017gyf, Chang:2018iay, Thorngren:2019iar, Koide:2021zxj, Choi:2021kmx, Kaidi:2021xfk, Roumpedakis:2022aik, Choi:2022jqy, Cordova:2022ieu, Shao:2023gho}.
These generalized symmetries also admit anomalies and the corresponding matching constraints \cite{Gaiotto:2017yup, Thorngren:2019iar, Ji:2019jhk, Levin:2019ifu, Hayashi:2022fkw, Karasik:2022kkq, Apte:2022xtu, Mekareeya:2022spm, Choi:2023xjw, Zhang:2023wlu, Cordova:2023bja, Seiberg:2023cdc, Antinucci:2023ezl, Seifnashri:2023dpa, Copetti:2024rqj, Pace:2024acq, Putrov:2024uor, Seiberg:2024gek, Putrov:2025xmw, Antinucci:2025fjp}.
Most such constraints forbid the existence of a unique gapped ground state preserving the symmetry---an \emph{invertible} phase---forcing the theory to either be gapless, exhibit spontaneous symmetry breaking (SSB), or flow to a nontrivial topological quantum field theory (TQFT).
Moreover, recent work shows that certain anomalies of finite symmetries can exclude the TQFT option as well \cite{Cordova:2019bsd, Cordova:2019jqi, Apte:2022xtu, Brennan:2023ynm, Hsin:2025ria}.

For continuous symmetries in continuum theories, the mechanism is well-understood: if an anomaly forces SSB of a continuous symmetry, gaplessness follows from the Nambu--Goldstone theorem \cite{Nambu:1960tm, Goldstone:1961eq, Goldstone:1962es}.
For this reason, previous studies on \emph{symmetry enforced gaplessness}\footnote{In this paper, by ``symmetry enforced gaplessness'' we mean that the symmetry data excludes gapped phases even when spontaneous symmetry breaking is allowed.} have largely relied on anomalous continuous symmetries \cite{Wang:2014lca, Wang:2016gqj, Sodemann:2016mib, Wang:2017txt, Chatterjee:2024gje, Pace:2024oys, Kim:2025tzc} (see also \cite{Sahay:2025xdg} for a different mechanism of enforced gaplessness).
However, a systematic understanding of how to realize anomalous symmetries on the lattice, especially continuous symmetries, is still incomplete, despite significant recent progress \cite{Chatterjee:2024gje, Pace:2024oys, Gioia:2025bhl, Pace:2025rfu}.

In this paper, we develop a mechanism for enforced gaplessness that requires only discrete symmetries and non-anomalous continuous symmetries at the UV scale. Since both ingredients have well-understood lattice realizations, this provides a more concrete path to lattice realizations of enforced gaplessness.

The central idea is a compatibility constraint coming from multiple embeddings of the same symmetry: when a symmetry $\mathcal{E}$ is simultaneously embedded into two larger symmetries $\mathcal{C}$ and $\mathcal{D}$, any gapped phase must be consistent with \emph{both} embeddings---that is, it must arise as the restriction of some gapped $\mathcal{C}$-symmetric phase \emph{and} some gapped $\mathcal{D}$-symmetric phase.
When no such compatible gapped phase exists, the system is forced to be gapless.
We call such a configuration of embeddings a \textit{symmetry span}.

To state this precisely, consider global symmetries that fit into the following commutative diagram:
\begin{equation}
    \begin{tikzcd}
       \mathcal{E}\arrow[r, hook, "i_{\mathcal{C}}"] \arrow[d, hook', "i_{\mathcal{D}}"'] & \mathcal{C} \arrow[d, hook'] \\
      \mathcal{D} \arrow[r, hook] & \mathcal{S}^{\text{faith}} ,\\
    \end{tikzcd}
\end{equation}
Here, $\mathcal{E},\mathcal{C},\mathcal{D}$ are tensor categories describing global symmetries\footnote{For both cases of continuous symmetries and symmetries on the lattice, the precise mathematical formulation is not just a tensor category. For continuous symmetry, see \cite{Jia:2025vrj,Stockall:2025ngz} for mathematical formulations and e.g.~\cite{Brennan:2024fgj,Antinucci:2024zjp,Bonetti:2024cjk,Argurio:2024oym,Gagliano:2024off,Cvetic:2025kdn,Bonetti:2025dvm,Antinucci:2024bcm,Apruzzi:2025hvs,Jia:2025jmn} for SymTFT perspectives. In this paper we do not need the details of such descriptions.}, and $\mathcal{S}^{\text{faith}}$ is a larger symmetry that contains both $\mathcal{C}$ and $\mathcal{D}$ as sub-categories.
Physically, this means we have a Hilbert space carrying two sets of symmetry operators---those of $\mathcal{C}$ and those of $\mathcal{D}$---which share common operators forming the sub-symmetry $\mathcal{E}$.

Let us state our main result.
The criterion for gaplessness can be stated precisely in terms of the classification of gapped phases.
Let $\TQFT(\mathcal{E})$ denote the category of gapped theories (TQFTs) with $\mathcal{E}$ symmetry.
Via the embedding functors $i_{\mathcal{C}}$ and $i_{\mathcal{D}}$, every gapped theory with $\mathcal{C}$ or $\mathcal{D}$ symmetry restricts to a gapped theory with $\mathcal{E}$ symmetry.
This defines two sub-categories of $\TQFT(\mathcal{E})$: the pullbacks $i_{\mathcal{C}}^*\TQFT(\mathcal{C})$ and $i_{\mathcal{D}}^*\TQFT(\mathcal{D})$.

Crucially, any gapped phase with the full symmetry $\Sfaith$ must simultaneously be a valid $\mathcal{C}$-symmetric phase and a valid $\mathcal{D}$-symmetric phase.
Restricting to the common sub-symmetry $\mathcal{E}$, this phase must therefore lie in both pullback categories.
Thus, a necessary condition for gapped phases to exist is:
\begin{equation}\label{eq:gapped condition}
    i_{\mathcal{C}}^*\TQFT(\mathcal{C}) \cap i_{\mathcal{D}}^*\TQFT(\mathcal{D}) \neq \{0\}.
\end{equation}
When the intersection is empty, no gapped theory can realize both $\mathcal{C}$ and $\mathcal{D}$ symmetries, and the system is necessarily gapless---this is symmetry enforced gaplessness from a symmetry span.

While this condition is quite general and applicable in any dimension, in this paper we mainly focus on the following particular symmetry span in 1+1 dimensions:
\begin{equation}
    \begin{tikzcd}
      \Vect_{H} \arrow[r, hook, "i"] \arrow[d, hook', "i_{(\phi,\beta)}"'] & \mathcal{C} \arrow[d, hook'] \\
      \mathbf{Vect}_{G} \arrow[r, hook] & \mathcal{S}^{\text{faith}} .\\
    \end{tikzcd}
\end{equation}
Here, $\mathcal{C}$ is a fusion category of a non-invertible symmetry; $H,G$ are some groups without anomalies; and $\Vect_{H},\mathbf{Vect}_{G}$ are the corresponding fusion categories describing the symmetries.
We use the notation $\mathbf{Vect}_{G}$ to emphasize that $G$ might be a continuous group, and thus the precise mathematical formulation needs more care than just considering $G$-graded vector spaces \cite{Jia:2025vrj,Stockall:2025ngz}.
$i\colon \Vect_{H}\hookrightarrow \mathcal{C}$ is the inclusion functor and $i_{(\phi,\beta)}$ is the embedding functor specified by a group homomorphism $\phi\colon H\rightarrow G$ and a 2-cocycle $\beta\in \mathrm{H}^2(H,\U(1))$ (see the main text for details).

Our main goal is to formulate the general criterion \eqref{eq:gapped condition} that any gapped phase must satisfy in the presence of a symmetry span.
We also illustrate the criterion by providing explicit symmetry spans for which \eqref{eq:gapped condition} is violated.
In such cases, the IR theory cannot be described by a relativistic gapped phase with a TQFT description and a finite-dimensional ground state space.
We further study how these symmetries are realized in concrete physical systems, largely using conformal field theories (CFTs), and we also discuss lattice realizations of the relevant symmetry embeddings via explicit lattice Hamiltonians.

The gaplessness argument does not rely on anomalies of continuous symmetries, though all continuum examples we present flow to IR theories with anomalous continuous symmetries.
Nevertheless, one can find many lattice realizations of symmetry spans that agree with the gaplessness argument.
While our lattice constructions do not realize anomalous continuous groups in the UV, they can flow to gapless theories whose IR descriptions exhibit anomalous continuous symmetries.
From this perspective, the span mechanism provides a controlled route to lattice models that realize anomalous continuous symmetries emergently in the IR.

Finally, the $\mathcal{C}$ symmetry in the span need not be both non-invertible and anomalous---either property alone can suffice for the gaplessness argument to apply. We provide examples where $\mathcal{C}$ is invertible or non-anomalous, yet the span still enforces gaplessness.

This paper is organized as follows. 
In Sec.~\ref{sec:non-inv sym 1+1D}, we review non-invertible symmetries and the classification of gapped TQFTs in 1+1 dimensions. 
Readers who are familiar with these topics may skip this section.
In Sec.~\ref{sec:sym_embedding}, we discuss embeddings of global symmetries and their equivalences based on some examples. 
Then in Sec.~\ref{sec:enforced_gaplessness}, we state the gaplessness argument, which is the main result of this work.
In Sec.~\ref{sec:conti_example}, we provide some examples of gapless theories that agree with the argument.
In Sec.~\ref{sec:lattice_realization}, we construct explicit lattice Hamiltonians realizing the symmetry spans, including spin chains with $\TY(\mathbb{Z}_N)$ and $\Rep(D_8)$ embeddings, and demonstrate their gaplessness.
We conclude with an outlook in Sec.~\ref{sec:outlook}.
Appendix~\ref{sec:bosonization} collects the conventions for bosonization, fermionization, and the Kennedy--Tasaki transformation, and Appendix~\ref{sec:spin(4)} identifies the symmetry span in the $\mathrm{Spin}(4)_1$ WZW CFT via its lattice realization.

\section{Non-invertible symmetries in 1+1D}\label{sec:non-inv sym 1+1D}
Finite symmetries in 1+1 dimensions are described by (unitary) fusion categories \cite{Bhardwaj:2017xup, Chang:2018iay, Thorngren:2019iar}.
In this section, we briefly review this fusion-category description, the associated classification of gapped phases, and a basic example that will be important later.
Roughly speaking, fusion categories are generalizations of finite groups.
Fusion rules, (weakened) associativity conditions, and other data are required to specify a fusion category.
For instance, a non-anomalous finite group $G$ symmetry is described by the fusion category $\Vect_G$, the category of $G$-graded vector spaces.
If the $G$ symmetry has an 't Hooft anomaly, the corresponding fusion category is $\Vect_G^\alpha$, the category of $G$-graded vector spaces with a nontrivial associator specified by a three-cocycle $\alpha\in \mathrm{H}^3(G,\U(1))$, also referred to as the $F$-symbol.
The associator has the following physical meaning. Let $\{U_g\}_{g\in G}$ be topological symmetry operators for a $G$ symmetry. There are two channels for fusing three symmetry operators into one, related by the $F$-symbol $\alpha(g_1,g_2,g_3)$:
\begin{equation}
\begin{tikzpicture}[baseline=(current bounding box.center), scale=0.8,
  every node/.style={font=\small}]
  \pgfmathsetmacro{\x}{1.25}
  \begin{scope}[shift={(0,0)}]
    \coordinate (L1) at (-\x, 1.4);
    \coordinate (L2) at (0, 1.4);
    \coordinate (L3) at (\x, 1.4);
    \coordinate (I1) at (-\x/2, 0.7);   %
    \coordinate (R)  at (0, 0);    %
    \draw[thick] (L1) -- (I1);
    \draw[thick] (L2) -- (I1);
    \draw[thick] (I1) -- (R);
    \draw[thick] (L3) -- (R);
    \draw[thick] (R) -- (0, -0.5);
    \node[above] at (L1) {$U_{g_1}$};
    \node[above] at (L2) {$U_{g_2}$};
    \node[above] at (L3) {$U_{g_3}$};
    \node[anchor=north east, inner sep=2pt] at ($(I1)!0.5!(R)$) {{\scriptsize $U_{g_1 g_2}$}};
    \node[below] at (0, -0.5) {$U_{g_1 g_2 g_3}$};
  \end{scope}
  \node at (3.0, 0.35) {$=\;\alpha(g_1,g_2,g_3)$};
  \begin{scope}[shift={(5.7,0)}]
    \coordinate (L1) at (-\x, 1.4);
    \coordinate (L2) at (0, 1.4);
    \coordinate (L3) at (\x, 1.4);
    \coordinate (I2) at (\x/2, 0.7);     %
    \coordinate (R)  at (0, 0);    %
    \draw[thick] (L2) -- (I2);
    \draw[thick] (L3) -- (I2);
    \draw[thick] (L1) -- (R);
    \draw[thick] (I2) -- (R);
    \draw[thick] (R) -- (0, -0.5);
    \node[above] at (L1) {$U_{g_1}$};
    \node[above] at (L2) {$U_{g_2}$};
    \node[above] at (L3) {$U_{g_3}$};
    \node[anchor=north west, inner sep=2pt] at ($(I2)!0.5!(R)$) {{\scriptsize $U_{g_2 g_3}$}};
    \node[below] at (0, -0.5) {$U_{g_1 g_2 g_3}$};
  \end{scope}
\end{tikzpicture}.
\end{equation}

The fusion category $\Vect_{G}^\alpha$ is \emph{invertible} in the sense that all simple objects have inverses under fusion. Fusion categories in general can also contain \emph{non-invertible} simple objects, and can therefore describe non-invertible symmetries.
A typical example of non-invertible symmetries is the Tambara--Yamagami category \cite{TAMBARA1998692} associated with a finite Abelian group $A$, denoted by $\TY(A)$.\footnote{Precisely speaking, additional data are needed to specify the Tambara--Yamagami fusion category associated with a finite group, though we often suppress them in this paper.} 
The category $\TY(A)$ consists of two types of simple objects. One type is labeled by elements of $A$; the fusion rules of these objects obey the group multiplication law, and in particular each such object is invertible. The other is the non-invertible object $\mathcal{N}$.
The fusion rules of $\TY(A)$ are explicitly given by
\begin{gather}
    \begin{split}
      a\otimes b=ab,\quad a,b\in A,\\
      a\otimes\mathcal{N}=\mathcal{N}\otimes a=\mathcal{N},\\
      \mathcal{N}\otimes\mathcal{N}=\bigoplus_{a\in A}a.
    \end{split}
\end{gather}
We see that $\TY(A)$ contains an invertible $A$ symmetry. 
Consistency conditions for $\TY(A)$ further require that the 't Hooft anomaly of the $A$ symmetry be absent. 
The physical meaning of Tambara--Yamagami symmetries is that a system with a $\TY(A)$ symmetry is invariant under gauging the non-anomalous $A$ symmetry, in the sense of a Kramers--Wannier-type duality \cite{Kramers:1941kn,Frohlich:2004ef}.

For a given $\mathcal{C}$ fusion category symmetry, one can classify gapped theories with the symmetry, i.e., TQFTs with the symmetry. 
Symmetry operators should consistently act on such TQFTs and the actions define structures of $\mathcal{C}$-module categories.
In general, gapped theories with a unitary fusion category symmetry $\mathcal{C}$ are classified by its module categories \cite{Thorngren:2019iar, Komargodski:2020mxz, Huang:2021zvu, Inamura:2021szw}. 
Roughly speaking, this classification is a generalization of SSB patterns to non-invertible symmetries; for example, the number of simple objects in module categories counts the degeneracies of vacua (ground states).
Thus, if a symmetry $\mathcal{C}$ admits an invertible TQFT, it admits the module category with one simple object. 
Not all fusion categories satisfy such a condition and fusion categories that do not admit any invertible TQFT are said to be \emph{anomalous} \cite{Thorngren:2019iar}.

The classification of gapped theories with non-invertible symmetries can be stated more mathematically. Let $\mathcal{M}$ be a left $\mathcal{C}$-module category:
\begin{equation}
  \mathcal{C}\times\mathcal{M}\rightarrow \mathcal{M};\quad (x,m)\mapsto x\triangleright m .
\end{equation}
This is equivalent to giving the data of a tensor functor from $\mathcal{C}$ to the category of endofunctors $\End(\mathcal{M})$.
We say that $\mathcal{C}$ is non-anomalous if there exists a $\mathcal{C}$-module category that consists of a single simple object, namely $\Vect$.
Since $\End(\Vect)\cong \Vect$, such a non-anomalous fusion category $\mathcal{C}$ admits a tensor functor $\mathcal{C}\rightarrow \Vect$, which is called a fiber functor.

Let us see how this picture works in the case of ordinary non-anomalous invertible $G$-group symmetries corresponding to $\Vect_G$.
Fiber functors $\Vect_G\rightarrow \Vect$ are classified by the second group cohomology $\mathrm{H}^2(G,\U(1))$, whose elements are called symmetry-protected topological (SPT) phases \cite{Gu:2009dr, Pollmann:2009mhk, Pollmann:2009ryx, Chen:2010zpc, Schuch:2010, Chen:2011pg, Levin:2012yb}.
The choice of an SPT phase can be seen more explicitly as follows.
Let $g,h\in G$ be group elements and let $c$ be the unique simple object in $\Vect$.
An SPT phase $\omega\in \mathrm{H}^2(G,\U(1))$ is specified by the $\U(1)$-valued multiplication \footnote{Technically, this account requires $\mathcal{C}$ to be a strict monoidal representative in its equivalence class. This means we regard $g\otimes h$ and $gh$ as different, but isomorphic, objects.}
\begin{equation}
  \Hom_{\Vect_G}(g\otimes h,gh)
  \cong\mathbb{C} \xrightarrow{\times\omega(g,h)} \Hom_{\Vect}(c\otimes c,c)\cong \mathbb{C}.
\end{equation}
More general gapped theories with non-anomalous $G$ symmetry are classified by pairs $(H,\alpha)$, where $H$ is a subgroup of $G$ defined up to conjugation, specifying an unbroken symmetry, and $\alpha\in \mathrm{H}^2(H,\U(1))$ is an SPT phase for the unbroken $H$ symmetry.

Returning to the non-invertible example $\mathcal{C}=\TY(A)$, let us consider the case $A=\mathbb{Z}_N$.
Though the invertible $\mathbb{Z}_N$ group symmetry in $\TY(\mathbb{Z}_N)$ is non-anomalous, the full $\TY(\mathbb{Z}_N)$ symmetry is anomalous, i.e., it does not admit any unique gapped ground state.
Indeed, if one had a unique gapped vacuum with a $\TY(\mathbb{Z}_N)$ symmetry, its ground state in particular would preserve the invertible $\mathbb{Z}_N$ symmetry. 
However, such an invertible state cannot be self-dual under $\mathbb{Z}_N$ gauging and has to be mapped to the fully $\mathbb{Z}_N$-symmetry-broken state, contradicting the assumption of uniqueness of vacuum.
To summarize, the $\mathbb{Z}_N$ symmetry inside $\TY(\mathbb{Z}_N)$ is necessarily spontaneously broken in some gapped degenerate ground states.%
We will revisit this point in Sec.~\ref{subsec:ZN_into_TY}.

\section{Symmetry embedding and pullback of module categories}\label{sec:sym_embedding}
Here we discuss situations where symmetry operators of a (non-invertible) $\mathcal{C}_1$ symmetry are identified with those of a larger (non-invertible) $\mathcal{C}_2$ symmetry.
Mathematically, this is described by a tensor functor between two fusion categories $f\colon\mathcal{C}_1\rightarrow \mathcal{C}_2$.
As we will see, the monoidal data of the tensor functor $f$ play an important role, especially for lattice realizations of such symmetry embeddings.
In this situation, every gapped theory with $\mathcal{C}_2$ symmetry can generally be pulled back to a gapped theory with $\mathcal{C}_1$ symmetry.
Specifically, for a given module map $g\colon\mathcal{C}_2\rightarrow\End(\mathcal{M})$, one obtains the corresponding $\mathcal{C}_1$-module map by composing $g$ with $f$.
However, not every gapped theory with $\mathcal{C}_1$ symmetry can be obtained as the pullback of a gapped theory with $\mathcal{C}_2$ symmetry.

\subsection{Finite group symmetry}
\label{subsec:pullback_group_symmetry}
Let us first see the case where a finite group symmetry $H_1$ is embedded into another group symmetry $H_2$.
In the absence of anomalies, the situation is given by specifying a tensor functor
\begin{equation}
  i\colon\Vect_{H_1} \rightarrow \Vect_{H_2}.
  \label{eq:group_functor}
\end{equation}
For fixed $H_1$ and $H_2$, the functor $i$ is classified by a group homomorphism $\phi\colon H_1\rightarrow H_2$ and a 2-cocycle $\beta\in \mathrm{H}^2(H_1,\U(1))$ \cite[Section 2.6]{Etingof:2015tensor}.

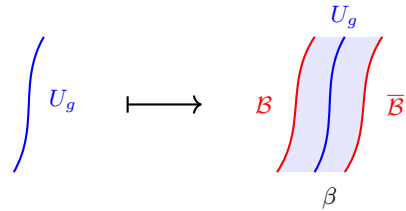
\begin{figure}[tbhp]
  \centering
  \begin{tikzpicture}[baseline=(current bounding box.center), scale=1.0,
    every node/.style={font=\small}]
    \begin{scope}[shift={(0,0)}]
      \draw[thick, blue] (-0.2,-0.9) .. controls (0.15,-0.3) and (-0.15,0.3) .. (0.2,0.9);
      \node[blue, right] at (0.15,0.05) {$U_g$};
    \end{scope}
    \draw[thick, |->] (1.3,0) -- ++(1,0);
    \begin{scope}[shift={(4,0)}]
      \fill[blue!10]
        (-0.7,-0.9) .. controls (-0.3,-0.3) and (-0.6,0.3) .. (-0.2,0.9)
        -- (0.7,0.9) .. controls (0.3,0.3) and (0.6,-0.3) .. (0.2,-0.9)
        -- cycle;
      \draw[thick, red] (-0.7,-0.9) .. controls (-0.3,-0.3) and (-0.6,0.3) .. (-0.2,0.9);
      \draw[thick, red] (0.2,-0.9) .. controls (0.6,-0.3) and (0.3,0.3) .. (0.7,0.9);
      \draw[thick, blue] (-0.2,-0.9) .. controls (0.15,-0.3) and (-0.15,0.3) .. (0.2,0.9);
      \node[blue, above] at (0.2,0.9) {$U_g$};
      \node[red, left] at (-0.65,0) {$\mathcal{B}$};
      \node[red, right] at (0.65,0) {$\overline{\mathcal{B}}$};
      \node at (0,-1.25) {$\beta$};
    \end{scope}
  \end{tikzpicture}
  \caption{Auto-equivalence of $\Vect_H$ by an SPT phase $\beta\in \mathrm{H}^2(H,\U(1))$. A symmetry operator $U_g$ is dressed by a strip of the SPT phase $\beta$, whose boundaries are the symmetry-breaking boundary $\mathcal{B}$ and its orientation reversal $\overline{\mathcal{B}}$.}
  \label{fig:auto_equiv_SPT}
\end{figure}

Taking $H_1=H_2$, this classification means there are nontrivial auto-equivalences of $\Vect_{H}$ specified by nontrivial 2-cocycles in $\mathrm{H}^2(H,\U(1))$, even if we consider the identity group homomorphism $\phi=\mathrm{id}$.
Physically, this auto-equivalence corresponds to dressing symmetry operators by a strip of a SPT phase $\beta \in \mathrm{H}^2(H,\U(1))$, see Fig.~\ref{fig:auto_equiv_SPT}.
Here, the strip of the SPT phase has a symmetry-breaking boundary condition $\mathcal{B}$ and its orientation reversal $\overline{\mathcal{B}}$, and the symmetry operator runs through the strip. 
(See also \cite{Kobayashi:2025ykb} for a related discussion on autoequivalences of categories in the context of anyons.)
A general functor \eqref{eq:group_functor} can be understood as first dressing symmetry operators by the SPT phase $\beta$ and then embedding them into the larger group $H_2$ through the homomorphism $\phi$.

To illustrate this functor concretely, let us consider the case where $H_1=\mathbb{Z}_2\times\mathbb{Z}_2$ and $H_2=\mathbb{Z}_4\times\mathbb{Z}_4$. We label the two $\mathbb{Z}_2$'s and the two $\mathbb{Z}_4$'s by $\mathbb{Z}_2^A\times\mathbb{Z}_2^B$ and $\mathbb{Z}_4^A\times\mathbb{Z}_4^B$, respectively.
At the level of objects, we fix the homomorphism $\phi$ as follows:
\begin{equation}\label{eq:hom_Z2Z2_Z4Z4}
  \phi\colon (a,b)\mapsto (2a,2b) \mod 4,\quad a,b\in \mathbb{Z}_2=\{0,1\}.
\end{equation}
To complete the definition of the tensor functor, we also need to specify the map between morphism spaces. Here we consider the following two choices:
\begin{align}
  \begin{split}
    &\Hom((a_1,b_1)\otimes (a_2,b_2),(a_1+a_2,b_1+b_2))\\
    &\xrightarrow{\times\omega_s} \Hom(i(a_1,b_1)\otimes i(a_2,b_2),i(a_1+a_2,b_1+b_2)),
  \end{split}
\end{align}
where
\begin{equation}
  \omega_s((a_1,b_1),(a_2,b_2))=(-1)^{sa_1 b_2}, \quad s=0,1.
\end{equation}

Let us see how SPT phases are related through the pullback map along \eqref{eq:group_functor} $i_{(\phi,\beta)}$ characterized by $\phi$ and $\beta$. We note that there are two (four) distinct SPT phases with $\mathbb{Z}_2\times\mathbb{Z}_2$ ($\mathbb{Z}_4\times\mathbb{Z}_4$) symmetry, since $\mathrm{H}^2(\mathbb{Z}_N\times\mathbb{Z}_N,\U(1))\cong \mathbb{Z}_N$. 
With the choice $\beta=0$, 
the pullback map $i_{(\phi,0)}^*$ acts on SPT phases as the pullback along the group homomorphism $\phi$.
\begin{equation}
  \phi^*\colon \mathrm{H}^2(\mathbb{Z}_4\times\mathbb{Z}_4,\U(1))\rightarrow \mathrm{H}^2(\mathbb{Z}_2\times\mathbb{Z}_2,\U(1)).
\end{equation}
In particular, the trivial SPT with $\mathbb{Z}_4\times\mathbb{Z}_4$ symmetry is pulled back to the trivial SPT with $\mathbb{Z}_2\times\mathbb{Z}_2$ symmetry. 
On the other hand, for the choice $s=1$, i.e.\ for the nontrivial $\beta \in \mathrm{H}^2(\mathbb{Z}_2\times\mathbb{Z}_2,\U(1))$, the pullback along $i_{(\phi,\beta)}$ is given by the composition of the pullback along $\phi$ and the addition of $\beta$:
\begin{equation}
  i_{(\phi,\beta)}^*\colon \mathrm{H}^2(H_2,\U(1))\xrightarrow{\phi^*} \mathrm{H}^2(H_1,\U(1)) \xrightarrow{+ \beta} \mathrm{H}^2(H_1,\U(1)).
  \label{eq:pullback_SPT_affine}
\end{equation}
In particular, the trivial SPT is sent to the nontrivial SPT, and the pullback map $i_{(\phi,\beta)}^*$ is not a group homomorphism between the cohomology groups. It is rather an affine map, or a map between $\mathrm{H}^2(H_2,\U(1))$-torsors.

More generally, a gapped phase with $H$ (more precisely, a simple module category over $\Vect_{H}$) is specified by a pair $(K,\gamma)$, where $K\subset H$ is a subgroup defined up to conjugation and $\gamma\in \mathrm{H}^2(K,\U(1))$ is an SPT phase for the unbroken $K$ symmetry \cite[Section~7.4]{Etingof:2015tensor}, \cite{Ostrik:2002ohv}.
We let the corresponding module category be denoted by $\Vect_{H/K}^\gamma$, and omit the subscript for SPTs where $H/K = \{\mathrm{pt}\}$.
The whole 2-category of $\Vect_{H}$-module categories $\Mod(\Vect_{H})$ is spanned by such module categories, i.e.\ a general module category looks like
\begin{equation}
  \bigoplus_{i} \Vect_{H/K_i}^{\gamma_i} \in \Mod(\Vect_{H}).
\end{equation}
Physically, the non-simple module corresponds to degenerate vacua not accounted for by spontaneous symmetry breaking of $H$ alone.
Note that the pullback of a simple module category is not necessarily simple, because we may forget some information of the broken part of the symmetry when we restrict the symmetry from $H_2$ to $H_1$.

For example, let us consider the case where $H_1=\mathbb{Z}_2$ and $H_2=\mathbb{Z}_4$ with the nontrivial embedding, and take the module category $\Vect_{\mathbb{Z}_4/\mathbb{Z}_2}$ corresponding to the phase where the $\mathbb{Z}_4$ symmetry is spontaneously broken to $\mathbb{Z}_2$, which happens to be $H_1$.
There are two vacua, both of which preserve the $H_1=\mathbb{Z}_2$ symmetry, and thus we have
\begin{equation}
  i^* \Vect_{\mathbb{Z}_4/\mathbb{Z}_2} \cong \Vect \oplus \Vect \in \Mod(\Vect_{\mathbb{Z}_2}).
\end{equation}

\subsection{Anomalous finite group symmetry}
\label{subsec:anomalous_finite_group}
Let us consider an anomalous finite group symmetry $\Vect_H^\omega$, where $\omega \in \mathrm{Z}^3(BH,\U(1))$ represents the 't Hooft anomaly.
A simple module category of $\Vect_H^\omega$ is labeled by a pair $(K,\gamma)$, where $K \subset H$ is a subgroup and $\gamma \in \mathrm{C}^2(BK,\U(1))$ is a trivialization of the pullback $\omega|_K$, i.e., $\delta\gamma = \omega|_K$.\footnotemark~
When $[\omega|_K] \neq 0$ in $\mathrm{H}^3(BK,\U(1))$, no such trivialization exists and $K$ is not allowed as an unbroken subgroup.

\footnotetext{
We note that not every distinct pair $(K,\gamma)$ gives rise to a distinct (inequivalent) module category over $\mathrm{Vec}_H^\omega$ \cite{natale2017equivalence}. 
}

As an example, let us consider $H_2 = H_1 \times \mathbb{Z}_N$ with the anomaly
\begin{equation}
  \omega = a \cup \alpha \in \mathrm{Z}^3(B(H_1 \times \mathbb{Z}_N), \U(1)),
\end{equation}
where $a \in \mathrm{H}^1(B\mathbb{Z}_N, \U(1))$ is the fundamental class and $\alpha \in \mathrm{H}^2(BH_1, \U(1))$.
We assume that $N\alpha = 0$ in $\mathrm{H}^2(BH_1, \U(1))$.
The subgroup $H_1 \subset H_1 \times \mathbb{Z}_N$ (embedded as $H_1 \times \{e\}$) satisfies $\omega|_{H_1} = 0$ since $a|_{H_1} = 0$.
Thus, $\Vect_{H_2/H_1}^\gamma$ with $\gamma \in \mathrm{H}^2(BH_1, \U(1))$ is a valid module category of $\Vect_{H_2}^\omega$.

Now consider the pullback of $\Vect_{H_2/H_1}^\gamma$ along the embedding $i \colon H_1 \to H_2 = H_1 \times \mathbb{Z}_N$ given by $h \mapsto (h, e)$.
The coset space $H_2/H_1 \cong \mathbb{Z}_N$ has $N$ vacua, each of which preserves the full $H_1$ symmetry under the embedding.
Due to the mixed anomaly $\omega = a \cup \alpha$, the SPT phase shifts by $\alpha$ as we move between adjacent vacua.
Explicitly, we have
\begin{equation}
  i^* \Vect_{H_2/H_1}^\gamma \cong \bigoplus_{k=0}^{N-1} \Vect^{\gamma + k\alpha} \in \Mod(\Vect_{H_1}).
\end{equation}
The $N$ different SPT phases $\gamma, \gamma + \alpha, \ldots, \gamma + (N-1)\alpha$ arise from the interplay between the original SPT $\gamma$ and the mixed anomaly.

\subsection{Embedding into a continuous symmetry}
Let us consider a finite group symmetry $H$ embedded into a continuous group symmetry $G$.
In terms of categories, we write
\begin{equation}
  i_{(\phi,\beta)} \colon \Vect_H \rightarrow \bfVect_G,
\end{equation}
where $\bfVect_G$ is the category of $G$-graded vector spaces with appropriate structures to capture the continuous nature of $G$ \cite{Jia:2025vrj,Stockall:2025ngz}.
Here, we assume the same classification of the functor $i_{(\phi,\beta)}$ by a group homomorphism $\phi\colon H\rightarrow G$ and a 2-cocycle $\beta\in \mathrm{H}^2(H,\U(1))$.

For continuous $G$, the Nambu--Goldstone theorem \cite{Nambu:1960tm, Goldstone:1961eq, Goldstone:1962es} says that a spontaneously broken continuous symmetry leads to gapless excitations.\footnote{
  In 1+1D, the Mermin--Wagner--Coleman theorem \cite{Mermin:1966fe, Hohenberg:1967zz, Coleman:1973ci} also forbids spontaneous breaking of continuous symmetries in a relativistic theory. However, in non-relativistic systems, spontaneous breaking of continuous symmetries is possible even in 1+1D, leading to gapless Nambu--Goldstone modes~\cite{Watanabe:2012hr,Hidaka:2012ym}. In any case, the point here is that there is no gapped phase with spontaneously broken continuous symmetry.
}
Thus, the (continuous) 2-category of $G$-symmetric TQFTs $\bfTQFT(\bfVect_G)$ only consists of the $G$-SPTs.\footnote{
  We avoid the term ``module category'' here since the proper definition of module categories over continuous fusion categories is not yet established and besides the point of this discussion. 
  For a finite symmetry $\mathcal{C}$, we use the terms ``$\mathcal{C}$-module category'' and ``$\mathcal{C}$-symmetric TQFT'' interchangeably.
}
The deformation class of bosonic 1+1d $G$-SPTs is classified by $\mathrm{H}^3(BG,\mathbb{Z})$ which is trivial for connected and simply connected Lie groups such as $G=\SU(N)$, and also for $G=\U(1)$ or a product thereof. For the latter, there is a nontrivial $S^1$ family of SPTs, but they are all deformation equivalent.
Any member of such a unique deformation class of $G$-SPTs is pulled back to the same deformation class of SPT phase with $H$ symmetry specified by $\beta\in \mathrm{H}^2(H,\U(1))$. 
Thus, for such cases, we have
\begin{multline}
  i^*_{(\phi,\beta)}(\bfTQFT(\bfVect_G)) \cong \\
  \{(\Vect^\beta)^{\oplus n}\}_{n\in \mathbb{Z}_{\geq 0}} \subsetneq \Mod(\Vect_H).
  \label{eq:pullback_continuous_only_SPT}
\end{multline}

\subsection{\texorpdfstring{$\!\mathbb{Z}_N$}{ZN} embedded into Tambara--Yamagami category}
\label{subsec:ZN_into_TY}
We can extend the above discussion to the case where a finite group symmetry $H$ is embedded into a non-invertible symmetry $\mathcal{C}$.
Let us consider the case $\mathcal{C}=\TY(\mathbb{Z}_p)$ for a prime $p$ and $H=\mathbb{Z}_p$.
We take the functor $i$ to map the simple objects in $\Vect_{\mathbb{Z}_p}$ to those in $\TY(\mathbb{Z}_p)$ labeled by the same group elements.

As noted in Sec.~\ref{sec:non-inv sym 1+1D},  
a Tambara--Yamagami symmetry $\TY(\mathbb{Z}_N)$ does not admit any gapped phase with a unique vacuum \cite{TAMBARA1998692,Thorngren:2019iar}.

 Furthermore, when $N$ is prime $N=p$, there is only one simple module category of $\TY(\mathbb{Z}_p)$, which is $\TY(\mathbb{Z}_p)$ itself, corresponding to the spontaneously broken phase of the $\TY(\mathbb{Z}_p)$ symmetry.
Pulling back this module category along the functor $i$ gives
\begin{equation}
  i^* \TY(\mathbb{Z}_p) \cong \Vect_{\mathbb{Z}_p} \oplus \Vect \in \Mod(\Vect_{\mathbb{Z}_p}),
  \label{eq:pullback_TY_to_Zp}
\end{equation}
where the first (second) summand corresponds to the vacuum that breaks (preserves) the $\mathbb{Z}_p$ symmetry. The duality defect $\mathcal{N}$ exchanges these two sets of vacua.
In particular, the pulled-back module category does not contain the unique simple module category $\Vect$ of $\Vect_{\mathbb{Z}_p}$, corresponding to the $\mathbb{Z}_p$-preserving gapped phase, and thus
\begin{multline}
  i^* \Mod(\TY(\mathbb{Z}_p)) \cong \\ \{(\Vect\oplus \Vect_{\mathbb{Z}_p})^{\oplus n}\}_{n\in\mathbb{Z}_{\ge 0}} \subsetneq \Mod(\Vect_{\mathbb{Z}_p}).
\end{multline}

For non-prime $N$, there are other simple modules. For example, there is one whose image under the pullback map $i^*$ is spanned by objects of the form
\begin{equation}
  \Vect_{\mathbb{Z}_N/\mathbb{Z}_{k_1}} \oplus \Vect_{\mathbb{Z}_N/\mathbb{Z}_{k_2}}
\end{equation}
with $k_1k_2 = N$.
When $N=M^2$ is a perfect square, there is also a simple module whose image under $i^*$ is $\Vect_{\mathbb{Z}_N/\mathbb{Z}_M}$, corresponding to a gapped phase that preserves the $\mathbb{Z}_M$ subgroup of the $\mathbb{Z}_N$ symmetry, which is self-dual under the $\mathbb{Z}_N$ gauging.

\subsection{\texorpdfstring{$\Rep(D_8)$}{Rep(D8)} symmetry and its pullback to \texorpdfstring{$\mathbb{Z}_2\times\mathbb{Z}_2$}{Z2xZ2}}
\label{subsec:RepD8_pullback}

Another notable example can be found in the case where $H=\mathbb{Z}_2\times\mathbb{Z}_2$ and $\mathcal{C}=\Rep(D_8)$.
Here, $D_8$ is the dihedral group of order eight, and $\Rep(D_8)$ is the representation category of $D_8$.
$D_8$ has four one-dimensional irreducible representations and one two-dimensional irreducible representation, and the one-dimensional ones have $\mathbb{Z}_2\times\mathbb{Z}_2$ fusion rules.
Thus, we take the embedding functor $i\colon \Vect_{\mathbb{Z}_2\times\mathbb{Z}_2}\rightarrow \Rep(D_8)$ to map the simple objects in $\Vect_{\mathbb{Z}_2\times\mathbb{Z}_2}$ to the one-dimensional irreducible representations of $D_8$.
Equivalently, it is known that $\Rep(D_8)$ can be realized as the Tambara--Yamagami category $\TY(\mathbb{Z}_2\times\mathbb{Z}_2)$ with an appropriate choice of $F$-symbols, and 
the above functor is equivalent to the natural embedding $\Vect_{\mathbb{Z}_2\times\mathbb{Z}_2}\rightarrow \TY(\mathbb{Z}_2\times\mathbb{Z}_2)$.

It is known that $\Rep(D_8)$ admits three distinct fiber functors, and in particular is non-anomalous \cite{TAMBARA1998692, Thorngren:2019iar}.
How are the three SPT phases with $\Rep(D_8)$ symmetry pulled back to those with $\mathbb{Z}_2\times\mathbb{Z}_2$ symmetry?
According to the classification of fiber functors of $\Rep(D_8)$ (see, e.g., \cite{Tambara:2000qlz}), all three SPT phases with $\Rep(D_8)$ symmetry are pulled back to the nontrivial SPT phase with $\mathbb{Z}_2\times\mathbb{Z}_2$ symmetry.

However, we can also define another embedding functor by precomposing the autoequivalence of $\Vect_{\mathbb{Z}_2\times\mathbb{Z}_2}$ corresponding to the nontrivial SPT phase with the natural embedding functor.
With this twisted embedding functor, all of the three SPT phases with $\Rep(D_8)$ symmetry are pulled back to the trivial SPT phase with $\mathbb{Z}_2\times\mathbb{Z}_2$ symmetry.

Physically, the distinction among the possible SPT phases obtained as pullbacks of the $\Rep(D_8)$ symmetry comes from the realization of the duality map \cite{Ando:2024hun}. The duality map corresponding to the natural embedding is the untwisted gauging of the $\mathbb{Z}_2\times\mathbb{Z}_2$ symmetry, while the duality map corresponding to the twisted embedding is the twisted gauging of $\mathbb{Z}_2\times\mathbb{Z}_2$ with the nontrivial SPT phase, which is known as the Kennedy--Tasaki transformation in the condensed matter literature \cite{kennedy1992hidden, Else:2013gsf, Duivenvoorden:2013tfa, Li:2023mmw}.

\section{Symmetry enforced gaplessness}\label{sec:enforced_gaplessness}
{\itshape Symmetry enforced gaplessness---Let $H,G$ be groups and let $\mathcal{C}$ be a fusion category.
Here $G$ may be a continuous group, and we denote the corresponding tensor category by $\mathbf{Vect}_{G}$.
We consider 1+1-dimensional systems with a global symmetry $\mathcal{S}^{\mathrm{faith}}$ that fit into the following commutative diagram of tensor functors:
\begin{equation}
  \begin{tikzcd}
    \Vect_{H} \arrow[r, hook, "i_{\mathcal{C}}"] \arrow[d, hook', "i_{(\phi,\beta)}"'] & \mathcal{C} \arrow[d, hook'] \\
    \mathbf{Vect}_{G} \arrow[r, hook] & \mathcal{S}^{\mathrm{faith}} ,\\
  \end{tikzcd}
  \label{diagram}
\end{equation}
which we call a \emph{symmetry span}. Let us denote the category of TQFTs with $\mathcal{C}$ symmetry and the category of TQFTs with $\mathbf{Vect}_{G}$ symmetry by $\TQFT(\mathcal{C})$ and $\TQFT(\mathbf{Vect}_{G})$, respectively.

In this situation, the system admits gapped phases (TQFTs) only if
\begin{equation}
  i_{\mathcal{C}}^\ast\TQFT(\mathcal{C}) \cap i_{(\phi,\beta)}^\ast\TQFT(\mathbf{Vect}_{G})\ne \{0\}.
\end{equation}
In other words, if this condition does not hold, then the theory cannot be a relativistic gapped theory in the deep IR.}

When $G$ is connected and we ask for \emph{gapped} phases, the right leg is particularly rigid: by the discussion in Sec.~\ref{sec:sym_embedding} (continuous symmetries cannot be spontaneously broken in a gapped phase), any gapped $G$-symmetric theory is a $G$-SPT, and its pullback to $H$ is necessarily a direct sum of  $H$-SPT's. In particular, if $G$ is either connected and simply connected, or $G=\U(1)$, the pullback along $i_{(\phi,\beta)}$ is given by a direct sum of copies of the \emph{fixed} $H$-SPT phase $\beta\in \mathrm{H}^2(H,\U(1))$ as summarized in \eqref{eq:pullback_continuous_only_SPT}.

\paragraph*{Remark 1.}
While the above argument excludes the existence of (relativistic) TQFT phases, it cannot exclude the existence of gapped phases without a TQFT description, such as fracton phases, especially in dimensions higher than two.
Nonetheless, in this paper we somewhat sloppily refer to the absence of TQFT phases as ``gaplessness''.

\paragraph*{Remark 2.}
From the data $H,G,\mathcal{C}$ and the embeddings $i_{(\phi,\beta)}$ and $i_{\mathcal{C}}$, one can define a universal symmetry $\mathcal{S}^{\mathrm{pushout}}$ fitting into the diagram \eqref{diagram}, characterized by the usual pushout/universality property. In this language, demanding the existence of a symmetry span is equivalently phrased as demanding that the symmetry factors through $\mathcal{S}^{\mathrm{pushout}}$. However, $\mathcal{S}^{\mathrm{pushout}}$ is typically extremely large: even when both $\mathcal{C}$ and $G$ come from finite group symmetries, $\mathcal{S}^{\mathrm{pushout}}$ corresponds to an amalgamated free product and is generally an infinite discrete group generated by alternating words in the two groups. The symmetry that acts faithfully on the system can nevertheless be much smaller, since the action factors through a morphism $\mathcal{S}^{\mathrm{pushout}}\to \mathcal{S}^{\mathrm{faith}}$. Thus, while analyzing $\mathcal{S}^{\mathrm{pushout}}$ is equivalent to analyzing the span, directly constructing and working with the pushout symmetry is often inconvenient; in this paper we instead analyze the span itself.

\paragraph*{Remark 3.}
The gaplessness argument applies in any spacetime dimension, with non-invertible symmetries described by appropriate higher categories. In this paper, we focus on 1+1-dimensional systems.

As we explained in Sec.~\ref{sec:sym_embedding}, the images of the pullback functors $i_{\mathcal{C}}^*$ and $i_{(\phi,\beta)}^*$ depend on the \emph{monoidal} data of the embeddings.
In the group-theoretic case this dependence is already visible at the level of SPTs: the induced map on $\mathrm{H}^2$ is in general \emph{affine} rather than a homomorphism, cf.\ \eqref{eq:pullback_SPT_affine}.
Equivalently, dressing the embedding by an $H$-SPT can shift what is realized as the ``trivial'' $H$-SPT after pullback.
Physically, this is fixed once one specifies the microscopic realization of the full set of symmetry operators (including non-invertible ones).

\subsection{Examples}
Deferring examples of concrete physical theories to the next section, we give here examples of non-invertible symmetries $\mathcal{C}$ and group symmetries $H,G$ that satisfy the assumptions of the gaplessness argument.

\paragraph*{Example 1.}
The first example is $H=\mathbb{Z}_N$ and $\mathcal{C}=\TY(\mathbb{Z}_N)$. As evident from the discussion in Sec.~\ref{sec:non-inv sym 1+1D}, $\TY(\mathbb{Z}_N)$ does not admit any $\mathbb{Z}_N$-preserving gapped theories. Namely,
\begin{equation}
  \Mod(\Vect_{\mathbb{Z}_N})\ni \Vect^{\oplus n}\notin i^\ast\Mod(\TY(\mathbb{Z}_N)),
  \label{eq:TY_no_ZN_preserving}
\end{equation}
for any positive integer $n$.
Here $\Vect^{\oplus n}$ denotes the phase with $n$ vacua with a trivial $\mathbb{Z}_N$ action.

On the other hand, if the $\mathbb{Z}_N$ symmetry is identified with a subgroup of a connected continuous symmetry $G$, the pulled-back $\mathbb{Z}_N$ TQFT must be a direct sum of SPT's, as noted earlier in this section. 
Given that the only $\mathbb{Z}_N$-symmetric SPT is the trivial one $\Vect$, this means the pullback has to be $\Vect^{\oplus n}$, contradicting \eqref{eq:TY_no_ZN_preserving}, and thus the theory is necessarily gapless.

In this example, the total symmetry $\mathcal{S}^{\text{faith}}$ is not simply a product $G\times \TY(\mathbb{Z}_N)$ for non-anomalous $G$.
This is because the self-duality under $\mathbb{Z}_N$ gauging, which is part of the $\TY(\mathbb{Z}_N)$ symmetry, changes the $G$ symmetry into another symmetry $G/\mathbb{Z}_N\times\mathbb{Z}_N$ with a mixed anomaly, where the latter $\mathbb{Z}_N$ is the quantum symmetry.
In the case $G=\U(1)$, one scenario for $\mathcal{S}^{\text{faith}}$ is a combination of a $\U(1)^2$ symmetry, whose mixed anomaly is given by a level $k$ coprime to $N$, together with the self-duality under $\mathbb{Z}_N$ gauging.

\paragraph*{Fermionized statement.}
For bosonic systems with a non-anomalous $\mathbb{Z}_2$ symmetry, one can map these systems to fermionic systems with the fermion parity symmetry $\mathbb{Z}_2^f$ by a procedure referred to as \textit{fermionization} \cite{Karch:2019lnn, Ji:2019jhk}.
Since the fermionization procedure is just a kind of discrete gauging, the gaplessness statement also holds for fermionized systems, and here we discuss how it can be directly read in the fermionized context.

On the one hand, consider a bosonic system with $\TY(\mathbb{Z}_2)$ symmetry.
Fermionization (see also \cite{Inamura:2022lun, Bhardwaj:2024ydc} for fermionization of fusion category symmetries) maps this to a fermionic system with $\mathbb{Z}_2^f\times\mathbb{Z}_2^{\text{Maj}}$ symmetry, whose anomaly corresponds to a generator of the $\mathbb{Z}_8$ classification of fermionic $\mathbb{Z}_2$ symmetry anomalies.%
\footnote{This classification arises from the Spin bordism group $\Omega_{3}^{\Spin}(B\mathbb{Z}_2)\cong \mathbb{Z}_8$ \cite{Kapustin:2014dxa, Freed:2016rqq, Yonekura:2018ufj}.
Depending on the choice of $F$-symbols, there are two inequivalent $\TY(\mathbb{Z}_2)$ categories: the Ising category and the $\mathrm{SU}(2)_2$ category.
Upon fermionization, the former corresponds to $1$ or $7 \bmod 8$, while the latter corresponds to $3$ or $5 \bmod 8$.}
We thus have the embedding $i_1\colon\mathbb{Z}_2^f\hookrightarrow\mathbb{Z}_2^f\times\mathbb{Z}_2^{\text{Maj}}$.

On the other hand, suppose that the theory also has a non-anomalous connected continuous group symmetry $G$. We assume that $G$ has a central $\mathbb{Z}_2$ symmetry and consider fermionization with respect to this $\mathbb{Z}_2$ symmetry.
The symmetry of the fermionized system is $G^f$, a non-anomalous symmetry group that has the fermion parity symmetry $\mathbb{Z}_2^f$ as a subgroup \cite{Ando:2024nlk}.  
Now the symmetry span is written as follows:
\begin{equation}
  \begin{tikzcd}
    \mathbb{Z}_2^f \arrow[r, hook, "i_1"] \arrow[d, hook', "i_2"'] & \mathbb{Z}_2^f\times\mathbb{Z}_2^{\text{Maj}} \arrow[d, hook'] \\
    G^f \arrow[r, hook] & \mathcal{S}^{\text{faith},f} .\\
  \end{tikzcd}
\end{equation}
To find the possible gapped theories realized as the pullback by $i_1$, we note that the anomalous symmetry $\mathbb{Z}_2^{\text{Maj}}$ has to be spontaneously broken if the theory is gapped, because the fermion parity is always preserved in fermionic systems.
Furthermore, the generator of $\mathbb{Z}_2^{\text{Maj}}$ acts by stacking the Arf invertible theory (Arf SPT).
Hence, if the theory were gapped, we would have some pairs of vacua that differ by the Arf SPT.

However, when $G$ is $\U(1)$ or simply connected, the deformation class of fermionic invertible theories with $G^f$ symmetry is trivial.
Thus, the pullback along $i_2$ can only yield copies of a single $\mathbb{Z}_2^f$-SPT phase---either all trivial or all Arf SPT, but not a mixture of both.
This contradicts the requirement from the pullback along $i_1$, and therefore the theory must be gapless.

Note that the total symmetry again satisfies $\mathcal{S}^{\text{faith},f}\neq G^f\times\mathbb{Z}_2^{\text{Maj}}$, which can be seen by mapping the fermionic system back to the bosonic system by bosonization.

As a concrete theory, consider for example the case where $G=\U(1)$.
A gapless theory with this span is realized as the free Dirac CFT, which has the vector and axial $\U(1)$ symmetries.
Both $\U(1)$ symmetries are extended by the fermion parity symmetry, and one can take either of the two $\U(1)$ symmetries for the span $\mathbb{Z}_2^f\hookrightarrow \U(1)^f$.
Note that each $\U(1)$ is non-anomalous, even though they carry a mixed anomaly.

\paragraph*{Example 2.}
Another example arises for $H=\mathbb{Z}_2\times\mathbb{Z}_2$ and $\mathcal{C}=\Rep(D_8)$.
We take the functor $i_\mathcal{C}\colon \Vect_H \rightarrow \Rep(D_8)$ to be the natural embedding, identifying $\Rep(D_8)$ with $\TY(\mathbb{Z}_2\times\mathbb{Z}_2)$.
As explained in Sec.~\ref{subsec:RepD8_pullback}, with this embedding the allowed $\Rep(D_8)$-SPTs all pull back to a single $\mathbb{Z}_2\times\mathbb{Z}_2$ SPT---either the trivial or the nontrivial one, depending on the choice of the embedding functor.
Similarly, if we embed $\mathbb{Z}_2\times \mathbb{Z}_2$ into a group $G$ with only the trivial SPT (again, $\U(1)\times\U(1)$ is a canonical choice), the pullback along $i_{(\phi,\beta)}$ yields only the $\mathbb{Z}_2\times\mathbb{Z}_2$ SPT $\beta$.
Thus, if we choose $i_\mathcal{C}$ and $i_{(\phi,\beta)}$ such that the pulled-back SPTs do not match, the theory is necessarily gapless.

\paragraph*{Example 3.}
The gaplessness argument also applies when $\mathcal{C}$ is a finite anomalous invertible symmetry.
Consider an embedding of a non-anomalous $\mathbb{Z}_2^A\times\mathbb{Z}_2^B$ symmetry into a $\mathbb{Z}_2^A\times\mathbb{Z}_2^B\times\mathbb{Z}_2^C$ symmetry with a type III anomaly $\omega = a \cup \alpha$, where $a \in \mathrm{H}^1(B\mathbb{Z}_2^C, \U(1))$ is the fundamental class and $\alpha \in \mathrm{H}^2(B(\mathbb{Z}_2^A\times\mathbb{Z}_2^B), \U(1))$ is the nontrivial SPT.
Explicitly, $\omega$ is given by
\begin{equation}
  \omega((a_1,b_1,c_1),(a_2,b_2,c_2),(a_3,b_3,c_3))=(-1)^{a_1 b_2 c_3},
\end{equation}
for $(a_i,b_i,c_i)\in\mathbb{Z}_2^A\times\mathbb{Z}_2^B\times\mathbb{Z}_2^C$.
In addition, we consider an embedding of $\mathbb{Z}_2^A\times\mathbb{Z}_2^B$ into a non-anomalous $\U(1)\times\U(1)$ symmetry.

This is an instance of the setup in Sec.~\ref{subsec:anomalous_finite_group} with $H_1 = \mathbb{Z}_2^A\times\mathbb{Z}_2^B$ and $N=2$.
The anomaly forces the $\mathbb{Z}_2^C$ symmetry to be spontaneously broken, while $\mathbb{Z}_2^A\times\mathbb{Z}_2^B$ must remain unbroken to avoid Goldstone modes.
This corresponds to the module category $\Vect_{H_2/H_1}^\gamma$ with $\gamma \in \mathrm{H}^2(B(\mathbb{Z}_2^A\times\mathbb{Z}_2^B), \U(1))$, whose pullback to $H_1$ is
\begin{equation}
  i^* \Vect_{H_2/H_1}^\gamma \cong \Vect^\gamma \oplus \Vect^{\gamma + \alpha},
\end{equation}
containing both the trivial and nontrivial $\mathbb{Z}_2^A\times\mathbb{Z}_2^B$ SPT phases.
However, pullbacks of $\U(1)\times\U(1)$-symmetric phases yield only a single SPT type.
This is a contradiction, and therefore the theory is necessarily gapless.

The two examples above---embedding $\mathbb{Z}_2\times\mathbb{Z}_2$ into $\Rep(D_8)$ and into the type III anomalous $\mathbb{Z}_2^3$---are closely related.
Indeed, $\Rep(D_8)$ and the type III anomalous $\mathbb{Z}_2^3$ symmetry are related by gauging a $\mathbb{Z}_2\times\mathbb{Z}_2$ subgroup \cite{Seifnashri:2024dsd, Ando:2024hun}.
In particular, there exists a twisted gauging of $\mathbb{Z}_2\times\mathbb{Z}_2$ compatible with the $\U(1)\times\U(1)$ embedding, in the sense that the quantum $\mathbb{Z}_2\times\mathbb{Z}_2$ symmetry is also embedded in $\U(1)\times\U(1)$.
We will see this relation based on a concrete lattice realization in Sec.~\ref{sec:lattice_realization}.
These correspondences---$\TY(\mathbb{Z}_2)\leftrightsquigarrow \mathbb{Z}_2^f\times\mathbb{Z}_2^{\text{Maj}}$ and $\Rep(D_8)\leftrightsquigarrow \Vect_{\mathbb{Z}_2^3}^\omega$---are summarized in Table~\ref{tab:gapless examples}.
\begin{table}[tbp]
  \centering
\begin{tikzpicture}[remember picture]
  \node (lefttable) at (0,0)
    {\begin{tabular}{c|c}
     $H$ & $\mathcal{C}$ \\ \hline
     $\mathbb{Z}_2$ & $\TY(\mathbb{Z}_2)$ \\
     $\mathbb{Z}_2\times\mathbb{Z}_2$ & $\Rep(D_8)$
     \end{tabular}};
  \node (righttable) at (4.5,0)
    {\begin{tabular}{c|c}
    $H$ & $\mathcal{C}$ \\ \hline
    $\mathbb{Z}_2^f$ & $\mathbb{Z}_2^f\times\mathbb{Z}_2^{\text{Maj}}$ \\
    $\mathbb{Z}_2\times\mathbb{Z}_2$ & $\Vect_{\mathbb{Z}_2^3}^\omega$
     \end{tabular}};
  \draw[<->] (lefttable.east) -- node[above]{gauging} (righttable.west);
\end{tikzpicture}
\caption{Examples of symmetry-enforced gaplessness and their relations under gauging. The left and right columns in each table show the subgroup $H$ and the larger symmetry category $\mathcal{C}$, respectively.}
  \label{tab:gapless examples}
\end{table}

\section{Continuum Examples}\label{sec:conti_example}
Here we provide examples of gapless theories that agree with the gaplessness argument.
In all of these examples, the full symmetry $\mathcal{S}^{\text{faith}}$ contains a continuous symmetry with non-trivial anomaly, which by itself ensures the gaplessness of the theory.
We could not conclude whether this is always the case for continuum theories that satisfy the assumption of the gaplessness argument.
It would be interesting to find a counterexample where the gaplessness cannot be attributed to the anomaly of continuous symmetries alone, or to prove that such a counterexample does not exist.

\subsection{\texorpdfstring{$\U(1)_{2k}$}{U(1)2k} WZW CFT}
One simple example of non-invertible enlargement of continuous symmetries that satisfies the conditions of the gaplessness argument is the symmetry of the $\U(1)_{2k}$ Wess--Zumino--Witten conformal field theory (WZW CFT) ($k\geq 2$). 
This theory is realized as the compact boson model at a specific radius. %

In this example, we take the continuous group $G$ in the span to be the $\U(1)^S$ shift symmetry, and take the finite group $H$ to be its $\mathbb{Z}_k$ subgroup.
Combining with T-duality, this theory is self-dual under gauging $\mathbb{Z}_{k}$ subgroup of the $\U(1)^S$ shift symmetry \cite{Thorngren:2021yso}, resulting in the $\TY(\mathbb{Z}_k)$ non-invertible symmetry, completing the span required for the gaplessness argument.

In this example, as the $\TY(\mathbb{Z}_k)$ symmetry is realized using the T-duality, the shift symmetry $\U(1)^S$ and the non-invertible symmetry generate together the winding symmetry $\U(1)^W$.
Thus, the total symmetry $\mathcal{S}^{\text{faith}}$ generated by the span contains the continuous symmetry $\U(1)^S\times\U(1)^W$, with a mixed anomaly between the two $\U(1)$ symmetries. 

\subsection{\texorpdfstring{$T^2$}{T2} CFT}\label{subsec:T2_CFT}

Let us find examples of gapless theories that agree with the gaplessness argument based on the non-invertible symmetry $\Rep(D_8)$ and the type III anomaly of $\mathbb{Z}_2^3$ symmetry.

For the former, we take $(\U(1)_4)^2$ WZW CFT, which is equivalent to the compact boson model on a two-dimensional torus target space with a specific choice of the metric and the zero B-field.
This theory has the $\TY(\mathbb{Z}_2)^2$ symmetry coming from the self-duality of each $\U(1)_4$ WZW CFT.
As its subcategory, we have the $\operatorname{Rep}(H_8)$ symmetry consisting of the four invertible objects forming the $\mathbb{Z}_2\times\mathbb{Z}_2$ group and one non-invertible object $\mathcal{N}$ with quantum dimension two.
By replacing the non-invertible object $\mathcal{N}$ with the product of $\mathcal{N}$ and the $\mathbb{Z}_2$ symmetry swapping the two compact bosons, we obtain the $\Rep(D_8)$ symmetry \cite{Thorngren:2019iar}.
As noted above, the $\mathbb{Z}_2\times\mathbb{Z}_2$ symmetry is embedded into the $\U(1)^2$ shift symmetry of the torus CFT, and thus the theory agrees with the gaplessness argument.

Using the Kennedy--Tasaki transformation, this $\Rep(D_8)$ symmetric theory is mapped to another torus CFT with the type III anomaly of $\mathbb{Z}_2^3$ symmetry.
We will analyze this transformation in more detail in Appendix \ref{sec:spin(4)}, but let us here just state the result.

The obtained theory is the $\mathrm{Spin}(4)_1$ WZW CFT, which is the product of compact bosons at self-dual radius, due to the isomorphism $\mathrm{Spin}(4)\cong \mathrm{SU}(2)\times \mathrm{SU}(2)$.
This theory has shift and winding $\U(1)$ symmetries of each compact boson. Let us label an element of this $\U(1)^4$ symmetry as $(\theta_1,\theta_2,\phi_1,\phi_2)$, where $\theta_i$ and $\phi_i$ are parameters of the shift and winding symmetries of the $i$-th compact boson, respectively.
Then, the continuous symmetry $G$ in the span is the $\U(1)^2$ subgroup generated by $(\theta,\theta,0,0)$ and $(0,0,\phi,-\phi)$.
Note that this $\U(1)^2$ symmetry is non-anomalous.

The three $\mathbb{Z}_2$ symmetries in the type III anomalous $\mathbb{Z}_2^3$ symmetry are generated by $(\pi,\pi,0,0)$, $(0,0,\pi,\pi)$, and the charge conjugation $C_1$ of one of the compact bosons, acting on the $\U(1)^4$ symmetry as
\begin{equation}
  C_1:(\theta_1,\theta_2,\phi_1,\phi_2)\mapsto (-\theta_1,\theta_2,-\phi_1,\phi_2).
\end{equation}
To understand the type III anomaly, we use the fact that a single compact boson at the self-dual radius has a $\mathbb{Z}_2^3$ symmetry generated by the shift by $\pi$, the winding by $\pi$, and the charge conjugation, with a type III anomaly \cite{Thorngren:2021yso}.
Although in the single compact boson case the shift and winding $\mathbb{Z}_2$ symmetries have a type II anomaly, in our setup the type II anomalies cancel between the two compact bosons, leaving only the type III anomaly.

To connect to the 't~Hooft anomaly of the full theory, one can check that the continuous symmetry $G=\U(1)^2$ combined with the charge conjugation $C_1$ generates the whole $\U(1)^4$ symmetry, which has a mixed anomaly.

Here we remark that either for the $\Rep(D_8)$ case or the type III anomaly case, the symmetry span exists in any rectangular torus CFT with zero B-field, not just at the specific point where the theory is equivalent to $(\U(1)_4)^2$ or $\mathrm{Spin}(4)_1$ WZW CFT.
The specialty of these points is that these theories can be fermionized to free fermion theories, and thus easily realized and analyzed in lattice models as we will see in the next section. We could not find lattice models realizing the symmetry span at other points in the moduli space of torus CFTs.

\subsection{Commuting triple}
Another way of producing a gaplessness-enforcing symmetry span involving a type III anomaly is to consider commuting triples in Lie groups.

A commuting triple in a Lie group $G$ (see e.g.\ \cite{Witten:1997bs,Borel:1999bx,Kac:1999gw}) is a triple of elements $g_1,g_2,g_3\in G$ such that they commute with each other. Let us call such a commuting triple disconnected if it cannot be continuously deformed into the trivial triple.
A key property of disconnected commuting triples is that they can give rise to a flat connection on the three-torus $T^3$ with non-trivial Chern--Simons invariant.
For our purpose, we can utilize this to find a finite subsymmetry in a 1+1D WZW CFT that has the type III anomaly.

The primary example of disconnected commuting triples is in $\Spin(7)$ \cite[Appendix I]{Witten:1997bs}.
A disconnected commuting triple in $\SO(7)$, which uplifts to that in $\Spin(7)$, is given by
\begin{align}
  g_1 &= \mathrm{diag}(+1,+1,+1,-1,-1,-1,-1),\\
  g_2 &= \mathrm{diag}(+1,-1,-1,+1,-1,+1,-1),\\
  g_3 &= \mathrm{diag}(-1,+1,-1,+1,+1,-1,-1).
\end{align}
We fix uplifts of these elements to $\Spin(7)$ and also denote them by $g_1,g_2,g_3$.
All of them are order two elements (either in $\SO(7)$ or $\Spin(7)$) and thus generate a $\mathbb{Z}_2^3$ symmetry.

Let us for now consider the $\Spin(7)_k$ WZW CFT and its chiral (i.e.\ anomalous) $\Spin(7)_k$ symmetry and its $\mathbb{Z}_2^3$ subgroup generated by the above commuting triple.
The inflow action for the $\Spin(7)_k$ symmetry is $k$ times the unit Chern-Simons action for $\Spin(7)$. To know the anomaly restricted to the $\mathbb{Z}_2^3$ subgroup, we need to compute the Chern-Simons invariant of the flat connection $A_{g_1,g_2,g_3}$ on $T^3$ specified by the commuting triple.
This is indeed computed in \cite{Borel:1999bx}, and we have
\begin{equation}
  \int_{T^3} \operatorname{CS}(A_{g_1,g_2,g_3}) = \frac{1}{2} \mod \mathbb{Z}.
\end{equation}
Thus we have the type III anomaly of $\mathbb{Z}_2^3$ symmetry when $k$ is odd. There are no type I or II anomalies because whenever we restrict to any $\mathbb{Z}_2$ or $\mathbb{Z}_2\times\mathbb{Z}_2$ subgroup, the commuting triple becomes connected to the trivial triple, and thus the Chern-Simons invariant vanishes.

In order to apply the gaplessness argument, we need to elaborate slightly more.
The $\Spin(7)_k$ WZW CFT has the full $(\Spin(7)_L\times \Spin(7)_R)/\mathbb{Z}_2$ symmetry, where the diagonal $\Spin(7)/\mathbb{Z}_2$ subgroup is non-anomalous. We let $(g_L;g_R)$ denote an element of the full symmetry, where $g_L,g_R\in \Spin(7)$, and $(g;g)$ with $g\in \Spin(7)$ denotes an element of the diagonal subgroup.
We take this $\Spin(7)/\mathbb{Z}_2$ symmetry as the continuous group $G$ in the span.
Then, we consider the $\mathbb{Z}_2^3$ subgroup generated by $(g_1;g_1),(g_2;g_2),(1;g_3)$. The first two elements generate a $\mathbb{Z}_2^2$ subgroup of the diagonal $\Spin(7)/\mathbb{Z}_2$ symmetry, while the whole $\mathbb{Z}_2^3$ subgroup has the type III anomaly when $k$ is odd, satisfying the assumption of the gaplessness argument.

The same construction can be applied to other Lie groups having disconnected commuting triples, such as $\Spin(n)$ ($n\geq 7$), $E_6$, $E_7$, and $E_8$ \cite{Witten:1997bs,Borel:1999bx,Kac:1999gw}.

\section{Lattice realization}\label{sec:lattice_realization}
We have argued in Sec.~\ref{sec:enforced_gaplessness} that some symmetry spans can enforce gaplessness. In this section, we provide lattice realizations of such symmetry embeddings that agree with the gaplessness argument.

Before proceeding to specific examples, we briefly discuss the broader context of lattice realizations of generalized symmetries.
Many studies have elucidated various lattice realizations and extractions of finite (anomalous) symmetries, including higher-form and non-invertible symmetries \cite{Else:2014vma, Kawagoe:2019jbx, Aasen:2020jwb, Kawagoe:2021gqi, Garre-Rubio:2022uum, Seifnashri:2023dpa, Inamura:2023qzl, Seiberg:2024gek, Meng:2024nxx, Kobayashi:2024dqj, Kapustin:2025nju, Kawagoe:2025ldx, Tu:2025bqf, Shirley:2025yji, Tantivasadakarn:2025txn, Feng:2025qgg, Czajka:2025mme}.
On the other hand, it remains unclear whether anomalous continuous symmetries can be realized exactly on the lattice with finite-dimensional Hilbert spaces.
Recent work \cite{Chatterjee:2024gje, Pace:2024oys, Gioia:2025bhl, Pace:2025rfu} demonstrates the realization of some anomalous continuous symmetries on the lattice, and their constructions are consistent with the (informally) conjectured impossibility of such realizations in finite-dimensional Hilbert spaces.
The lattice models we present below circumvent this difficulty by using only non-anomalous continuous symmetries in the UV, while still achieving enforced gaplessness through the symmetry span mechanism.

\subsection{Embedding to \texorpdfstring{$\TY(\mathbb{Z}_N)$}{TY(ZN)} symmetry}
Let us consider a one-dimensional spin-$1/2$ chain and denote the Pauli matrices acting on site $j$ by $X_{j},Y_{j},Z_{j}$.
Consider a $\mathbb{Z}_2$ symmetry generated by
\begin{equation}
  U\coloneqq\prod_{j}X_{j}.
\end{equation}
Clearly this $\mathbb{Z}_2$ symmetry is non-anomalous. An embedding of this $\mathbb{Z}_2$ symmetry into a $\U(1)$ symmetry is given by
\begin{equation}\label{eq:U(1)}
  U(\theta)=\prod_{j}e^{\frac{i\theta}{2} (1-X_{j})},
\end{equation}
which contains the $\mathbb{Z}_2$ symmetry as the $\theta=\pi$ specialization, $U(\pi)=U$.

Now consider another embedding of this $\mathbb{Z}_2$ symmetry into the $\TY(\mathbb{Z}_2)$ non-invertible symmetry.
A concrete realization of the non-invertible symmetry operator $\mathsf{D}$ of $\TY(\mathbb{Z}_2)$ is given by the so-called Kramers--Wannier duality transformation \cite{Li:2023mmw, Chen:2023qst, Seiberg:2024gek}. Specifically, $\mathsf{D}$ acts on the Pauli operators as
\begin{equation}
  \mathsf{D}X_{j}=Z_{j}Z_{j+1}\mathsf{D},\quad \mathsf{D}Z_{j}Z_{j+1}=X_{j+1}\mathsf{D}.
\end{equation}
We note that the non-invertible operator $\mathsf{D}$ generates lattice translations in the $\mathbb{Z}_2$-symmetric sector.\footnote{For this reason, the lattice realization of $\TY(\mathbb{Z}_2)$ differs from its continuum counterpart and is not a fusion category in the strict sense; see \cite{Seiberg:2024gek, Seifnashri:2023dpa} for detailed discussions. A complete mathematical framework for such lattice symmetries has not yet been established, but we do not need it for the purposes of this paper.}
An example of a local Hamiltonian invariant under both the $\U(1)$ and $\mathsf{D}$ symmetries is given by
\begin{align}\label{eq:U(1)_4}
  \begin{split}
    H_{\U(1)_4}&=\sum_{j}\left(Y_{j}Z_{j+1}-Z_{j}Y_{j+1}\right)\\
    &=-i\sum_{j}\left(X_{j}Z_{j}Z_{j+1}+Z_{j}Z_{j+1}X_{j+1}\right).
  \end{split}
\end{align}
This Hamiltonian flows to the $\U(1)_4$ WZW CFT in the low-energy limit and is therefore gapless.

We note that the total symmetry algebra generated by $U(\theta)$ and $\mathsf{D}$ is complicated. For instance, one finds that any Hamiltonian that commutes with both $U(\theta)$ and $\mathsf{D}$ also commutes with another $\U(1)$ symmetry
\begin{equation}\label{eq:U(1)_DW}
  V(\theta)=\prod_{j}e^{\frac{i\theta}{4}(1-Z_{j}Z_{j+1})},
\end{equation}
due to the relation $\mathsf{D}\,U(\theta)=V(2\theta)\mathsf{D}$ (see also \cite{Vernier:2018han,Pace:2024oys}).

Though each of the symmetries \eqref{eq:U(1)} and \eqref{eq:U(1)_DW} generates a $\U(1)$ symmetry, they do not commute for generic values of $\theta$, and thus the two $\U(1)$ symmetries do not realize a $\U(1)^2$ symmetry.
In this sense, it is not clear that the two $\U(1)$ symmetries \eqref{eq:U(1)} and \eqref{eq:U(1)_DW} can be regarded as realizing a $\U(1)^2$ symmetry with a mixed anomaly on the lattice, even though a $\U(1)^2$ symmetry is recovered in the low-energy sector of some specific Hamiltonians such as the Hamiltonian \eqref{eq:U(1)_4}.

Whereas we have not established a way to characterize the ``anomaly'' of the total symmetry generated by $U(\theta)$ and $\mathsf{D}$, or of the subalgebra generated by $U(\theta)$ and $V(\theta)$, we can determine the anomaly of the $\mathbb{Z}_2\times\mathbb{Z}_2$ symmetry generated by $U=U(\pi)$ and $V=V(\pi)$ using the methods developed in, e.g., \cite{Else:2013gsf, Seifnashri:2023dpa}. We find that it carries a mixed anomaly, which agrees with taking the $\mathbb{Z}_2\times\mathbb{Z}_2$ subgroup of a $\U(1)^2$ symmetry with a level-one mixed anomaly.

In general, one can show that any $\U(1)$ symmetry with a local generator never carries an anomaly in one-dimensional lattice systems.
To see this, let us assume that we can put the $\U(1)$ symmetry on a finite chain of arbitrary system size.
Namely, we restrict ourselves to the following equation holding for any system size $L$:
\begin{equation}
  \mathcal{U}(\theta)=\prod_{j=1}^{L}e^{i\theta q_{j}}=e^{i\theta Q},\quad Q=\sum_{j=1}^{L}q_{j}.
\end{equation}
Then we see that any local charge operator $q_{j}$ commutes with any other local charge operator $q_{k}$.
Let us consider a $\mathbb{Z}_N$ subgroup generated by $\mathcal{U}(2\pi/N)$.
If the $\U(1)$ symmetry is anomalous, we can always find an integer $N$ such that the corresponding $\mathbb{Z}_N$ subgroup is anomalous.\footnote{Recall that the anomaly of a bosonic $\U(1)$ symmetry in 1+1D is classified by an integer $k$, which corresponds to the level of $\U(1)_{2k}$. Then we can take any $N$ that is coprime to $k$ to obtain an anomalous $\mathbb{Z}_N$ subgroup.}
However, any $\mathbb{Z}_N$ symmetry generated by locally commuting charge operators $q_{j}$ carries only a trivial anomaly.
This is a contradiction, and we conclude that the $\U(1)$ symmetry of the form above never carries any $\U(1)$ anomalies.

What does the total symmetry generated by $U(\theta)$ and $\mathsf{D}$ look like? As discussed in \cite{Vernier:2018han,Pace:2024oys}, the subalgebra of the total symmetry generated by $A_0=\sum_{j}X_{j}$ and $A_{1}=\sum_{j}Z_{j}Z_{j+1}$ forms the Onsager algebra \cite{Onsager:1944cr}, which obeys the following commutation relations:
\begin{align}\label{eq:onsager}
  \begin{split}
    [A_{l},A_{m}]&=4G_{l-m},\quad l\geq m,\\
    [G_{l},A_{m}]&=2A_{l+m}-2A_{m-l},\\
    [G_{l},G_{m}]&=0.
  \end{split}
\end{align}

To apply the gaplessness argument of Sec.~\ref{sec:enforced_gaplessness} on the lattice, we do not need a full classification of gapped phases with the lattice $\TY(\mathbb{Z}_2)$ symmetry. It suffices to assume that $\TY(\mathbb{Z}_2)$ does not admit a gapped phase in which the $\mathbb{Z}_2$ symmetry is fully preserved. In the continuum, the only $\mathbb{Z}_2$ symmetric gapped phase is the trivial phase, which is sent to $\mathbb{Z}_2$ broken phase by the Kramers--Wannier duality.
Therefore, for the above assumption to be violated, there must exist a non-trivial $\mathbb{Z}_2$-preserving gapped phase unique to lattice systems that is invariant under the Kramers--Wannier duality. We are currently not aware of any such example.
Combined with the constraint from the $\U(1)$ embedding---which forces any gapped phase to preserve $\mathbb{Z}_2$---this establishes gaplessness from the symmetry span.

Independently, one can also demonstrate gaplessness for the present lattice models by assuming only that the Hamiltonian commutes with $A_0$ and $A_1$, following the proof given in \cite{Chatterjee:2024gje, Pace:2024oys}. Here we give an outline of the proof. We first use the Jordan--Wigner transformation, which maps Pauli operators to fermionic Majorana operators. Specifically, the Jordan--Wigner transformation maps local operators as
\begin{equation}
  X_{j}\mapsto ia_{j}b_{j},\quad Z_{j}Z_{j+1}\mapsto ib_{j}a_{j+1},
\end{equation}
where $a_{j},b_{j}$ are Majorana operators acting on site $j$ that obey the anticommutation relations $\{a_{i},a_{j}\}=\{b_{i},b_{j}\}=2\delta_{ij}$ and $\{a_{i},b_{j}\}=0$.
Then one can show that any local term in the Hamiltonian commuting with both $A_0$ and $A_1$ can be written in the form $\prod_{n}a_{i_{n}}$ or $\prod_{n}b_{i_{n}}$.
Considering infinitesimal $\U(1)$ transformations, we can further show that local terms in the Hamiltonian have to be of the form $ia_{i}a_{j}+ib_{i}b_{j}$.
Since such a local Hamiltonian is free, one can explicitly diagonalize it and verify gaplessness.

Now let us consider the generalization of this realization to $\TY(\mathbb{Z}_N)$ symmetry with $N>2$. Specifically, we consider a one-dimensional chain with $N$-state spins on each site. We denote the generalized Pauli operators acting on site $i$ by $\hat{X}_{i},\hat{Z}_{i}$, which obey the relations
\begin{gather}
  [\hat{X}_{i},\hat{X}_{j}]=[\hat{Z}_{i},\hat{Z}_{j}]=0,\\
  \hat{X}_{i}\hat{Z}_{j}=\omega^{\delta_{ij}}\hat{Z}_{j}\hat{X}_{i},\quad \hat{Z}_{i}^N=\hat{X}_{i}^N=I_{N\times N},
\end{gather}
where $\omega=\exp(2\pi i/N)$.
Consider the following $\U(1)$ charge operator:
\begin{equation}
  Q_N=\sum_{i=1}^{L}Q_{N,i},\quad (Q_{N,i})_{ab}\coloneqq \delta_{ab}\,(a-1),
\end{equation}
where $a,b=1,\ldots,N$. Then $U(\theta)=\exp(i\theta Q_N)$ generates a $\U(1)$ symmetry and, in particular, contains the $\mathbb{Z}_N$ symmetry generated by $\prod_{i}\hat{Z}_{i}$ as a $\mathbb{Z}_{N}$ subgroup.
We can rewrite the local charge operator as
\begin{equation}
  Q_{N,i}=\frac{N-1}{2}-\sum_{n=1}^{N-1}\frac{\hat{Z}_{i}^n}{1-\omega^{-n}}.
\end{equation}
As in the case $N=2$, we have a concrete realization of an embedding of the $\mathbb{Z}_N$ symmetry into the $\TY(\mathbb{Z}_N)$ symmetry.
The non-invertible symmetry operator $\mathsf{D}_N$ of $\TY(\mathbb{Z}_N)$ acts on the generalized Pauli operators as
\begin{equation}
  \mathsf{D}_{N}\hat{Z}_{j}=\hat{X}_{j}^{\dagger}\hat{X}_{j+1}\,\mathsf{D}_N,\quad \mathsf{D}_N\hat{X}_{j}^{\dagger}\hat{X}_{j+1}=\hat{Z}_{j+1}\,\mathsf{D}_N.
\end{equation}
Then the gauging dual of the local $\U(1)$ charge, denoted by $\tilde{Q}_{N,i}$, is written as
\begin{equation}
  \tilde{Q}_{N,i}=\frac{N-1}{2}-\sum_{n=1}^{N-1}\frac{\hat{X}_{i}^{-n} \hat{X}_{i+1}^n}{1-\omega^{-n}}.
\end{equation}
It is known that when we define the following two operators
\begin{equation}
  A_0=\frac{4}{N}\sum_{i}\sum_{n=1}^{N-1}\frac{\hat{Z}_{i}^n}{1-\omega^{-n}},\quad A_1=\frac{4}{N}\sum_{i}\sum_{n=1}^{N-1}\frac{\hat{X}_{i}^{-n} \hat{X}_{i+1}^n}{1-\omega^{-n}},
\end{equation}
then they again generate the same Onsager algebra as in \eqref{eq:onsager} \cite{VONGEHLEN1985351}.
A concrete 2-local lattice model that commutes with both $Q_N$ and $\tilde{Q}_N$ is given in \cite{Vernier:2018han}.
They studied the model in detail and showed that it is gapless and has central charge $c=1$ or $c=3(N-1)/(N+1)$, depending on the sign of the overall Hamiltonian.
This model provides a lattice realization of the gaplessness-enforcing symmetry span for general $N$: the $\mathbb{Z}_N$ subgroup of the $\U(1)$ symmetry generated by $Q_N$ is simultaneously embedded into the $\TY(\mathbb{Z}_N)$ symmetry, and the gaplessness argument of Sec.~\ref{sec:enforced_gaplessness} applies just as in the $N=2$ case. Notably, for $N>2$ this lattice model is not dual to free fermions, making it a genuinely interacting example of a lattice Hamiltonian whose gaplessness is enforced by the symmetry span.

\subsection{\texorpdfstring{$\Rep(D_8)$}{Rep(D8)} symmetry and type III anomaly}
Let us consider a one-dimensional spin-$1/2$ chain with an even number of sites and a $\mathbb{Z}_2\times\mathbb{Z}_2$ symmetry generated by
\begin{equation}\label{eq:Z2Z2_sym}
  U_{o}=\prod_{j:\text{odd}}X_{j},\quad U_{e}=\prod_{j:\text{even}}X_{j}.
\end{equation}
We consider an embedding of this $\mathbb{Z}_2\times\mathbb{Z}_2$ symmetry into a $\U(1)\times\U(1)$ symmetry. We take the following two $\U(1)$ charges:
\begin{equation}\label{eq:U1U1_charge}
  Q_{o}=\frac{1}{2}\sum_{j:\text{odd}}(1-X_{j}),\quad Q_{e}=\frac{1}{2}\sum_{j:\text{even}}(1-X_{j}).
\end{equation}
Let us further consider an embedding of this $\mathbb{Z}_2\times\mathbb{Z}_2$ symmetry into the $\Rep(D_8)$ non-invertible symmetry.
A concrete realization of the non-invertible symmetry operator $\mathsf{S}$ of $\Rep(D_8)$ is given by the following action on the Pauli operators:
\begin{equation}
  \mathsf{S}X_{j}=Z_{j-1}Z_{j+1}\mathsf{S},\quad \mathsf{S}Z_{j-1}Z_{j+1}=X_{j}\mathsf{S}.
\end{equation}
One can see that any uniquely gapped Hamiltonian with this $\Rep(D_8)$ symmetry must be a nontrivial SPT Hamiltonian with respect to the $\mathbb{Z}_2\times\mathbb{Z}_2$ symmetry \eqref{eq:Z2Z2_sym}.
To see this, we note that the following cluster Hamiltonian has a unique gapped ground state and is invariant under the $\Rep(D_8)$ symmetry:
\begin{equation}
  H_{\text{cluster}}=-\sum_{j}X_{j-1}Z_{j}X_{j+1}.
\end{equation}
This Hamiltonian is known to be in a nontrivial SPT phase with respect to the $\mathbb{Z}_2\times\mathbb{Z}_2$ symmetry.
On the other hand, any fiber functor of $\Rep(D_8)$ is pulled back to the same fiber functor of $\mathbb{Z}_2\times\mathbb{Z}_2$ symmetry, as noted in Sec.~\ref{sec:sym_embedding}, which concludes the claim.
We note that this does not contradict the fact that $\Rep(D_8)$ is non-anomalous and admits a unique gapped ground state realized in an on-site fashion \cite{Meng:2024nxx}. Indeed, one can also define another embedding functor by twisting the map on morphisms using the multiplication of the nontrivial $\mathbb{Z}_2\times\mathbb{Z}_2$ SPT phase. The pullback of any of the three $\Rep(D_8)$ SPT phases through this twisted functor gives the trivial SPT phase with $\mathbb{Z}_2\times\mathbb{Z}_2$ symmetry.

Similar to the previous example, any Hamiltonian that commutes with $Q_{o}$, $Q_{e}$, and $\mathsf{S}$ also has to commute with the following two $\U(1)$ charges:
\begin{align}\label{eq:U1U1_charge_D8_pair}
  \begin{split}
    \widetilde{Q}_{o}&=\frac{1}{2}\sum_{j:\text{odd}}(1-Z_{j-1}Z_{j+1}),\\
    \widetilde{Q}_{e}&=\frac{1}{2}\sum_{j:\text{even}}(1-Z_{j-1}Z_{j+1}).
  \end{split}
\end{align}
Note that the two pairs $Q_o,\widetilde{Q}_e$ and $Q_e,\widetilde{Q}_o$ do not commute and generate complicated algebras.

Based on the discussion in the previous sections, we expect that this embedding agrees with the gaplessness argument. Indeed, we again obtain the Onsager algebra for odd and even sites separately as a subalgebra of the total symmetry.
Nevertheless, gaplessness may still be nontrivial, since the symmetry can allow couplings between Majorana operators on odd and even sites.
Fortunately, one can show that such couplings are also forbidden by the symmetry, as we now explain.

We again use the Jordan--Wigner transformation separately on the odd and even sites.
We note that we need an even number of Majorana operators both for odd sites and for even sites separately due to the fermion parity symmetry for both odd and even sites.
For odd $i,j$, the following sequence of actions maps $a_{i}b_{j}$ to $a_{i-2}b_{j+2}$:
\begin{align}
  \begin{split}
    a_{i}b_{j}&\xmapsto{\mathsf{S}}b_{i-1}a_{j+1}\xmapsto{U_{e}(\pi/2)}-a_{i-1}b_{j+1}\\
  &\xmapsto{\mathsf{S}}-b_{i-2}a_{j+2}\xmapsto{U_{o}(\pi/2)}a_{i-2}b_{j+2}.
  \end{split}
\end{align}
On the other hand, for even $i,j$, the same sequence of actions maps $a_{i}b_{j}$ to itself:
\begin{align}
  \begin{split}
    a_{i}b_{j}&\xmapsto{\mathsf{S}}b_{i-1}a_{j+1}\xmapsto{U_{e}(\pi/2)}b_{i-1}a_{j+1}\\
  &\xmapsto{\mathsf{S}}a_{i}b_{j}\xmapsto{U_{o}(\pi/2)}a_{i}b_{j}.
  \end{split}
\end{align}
From these calculations, we see that any local term consists of Majorana operators supported only on odd sites or only on even sites; namely, Majorana operators on odd and even sites cannot be coupled.
Similarly, one can find a symmetry action that maps $a_{i}b_{j}$ to $a_{i-2}b_{j+2}$ for even $i$ and to itself for odd $i$.
Hence, any coupling of Majorana operators such as $ia_{i}b_{j}$ is also forbidden by the symmetry.
Since we are now allowed to have only couplings of Majorana operators on odd sites or on even sites separately, the Hamiltonian is again free and gapless.

Let us now consider an embedding of the symmetry into both $\U(1)\times\U(1)$ and $\mathbb{Z}_2^3$ with the type III anomaly.
For the embedding into $\U(1)\times\U(1)$, we take the same $\U(1)$ charges $Q_{o},Q_{e}$ as in \eqref{eq:U1U1_charge}.
The other $\mathbb{Z}_2$ symmetry $U_{c}$, which carries the type III anomaly with the $\mathbb{Z}_2\times\mathbb{Z}_2$ symmetry generated by \eqref{eq:Z2Z2_sym}, acts on the Pauli operators as
\begin{equation}
  U_{c}X_{j}U_{c}^\dagger=Z_{j-1}X_{j}Z_{j+1},\quad U_{c}Z_{j}U_{c}^\dagger=Z_{j}.
\end{equation}
This symmetry $U_c$ is realized as a unitary operator, which is discussed in \cite{Tantivasadakarn:2021wdv, Li:2022nwa, Seifnashri:2024dsd}.\footnote{There are several ways to write this operator, but its action on the Pauli operators is the same, and they carry the same type III anomaly.}
Similar to the previous $D_8$ case, the $\mathbb{Z}_2$ symmetry $U_c$ also gives another pair of $\U(1)$ charges:
\begin{align}\label{eq:U1U1_charge_pair}
  \begin{split}
    \widetilde{Q}_{o}&=\frac{1}{2}\sum_{j:\text{odd}}(1-Z_{j-1}X_{j}Z_{j+1}),\\
    \widetilde{Q}_{e}&=\frac{1}{2}\sum_{j:\text{even}}(1-Z_{j-1}X_{j}Z_{j+1}).
  \end{split}
\end{align}
According to the discussion in the previous sections, we expect that any Hamiltonian that commutes with $Q_{o}$, $Q_{e}$, and $U_{c}$ is gapless. Here we demonstrate this using the non-invertible duality transformation that maps the $\mathbb{Z}_2^3$ symmetry with the type III anomaly to the $\Rep(D_8)$ symmetry \cite{Seifnashri:2024dsd}.
The non-invertible duality transformation, denoted by $\mathsf{KT}$, is realized as the so-called Kennedy--Tasaki transformation \cite{kennedy1992hidden, Else:2013gsf, Duivenvoorden:2013tfa, Li:2023mmw}, which acts on the Pauli operators as
\begin{equation}
  \mathsf{KT}X_{j}=X_{j}\mathsf{KT},\quad \mathsf{KT}Z_{j-1}Z_{j+1}=Z_{j-1}X_{j}Z_{j+1}\mathsf{KT}.
\end{equation}
Using this transformation, one can map the $\mathbb{Z}_2^3$ symmetry generated by $U_{o},U_{e},U_{c}$ to the $\Rep(D_8)$ symmetry generated by $U_{o},U_{e},\mathsf{S}$, and the two $\U(1)$ charges \eqref{eq:U1U1_charge} are mapped to the same $\U(1)$ charges under the transformation.
Though the Kennedy--Tasaki transformation is non-invertible, it just exchanges some symmetry sectors. Indeed, one can define the KT transformation as a unitary transformation for open boundary conditions.
Assuming that the existence of a gap in a many-body Hamiltonian does not rely on the details of the boundary conditions, we conclude that any Hamiltonian that commutes with $Q_{o}$, $Q_{e}$, and $U_{c}$ is also gapless.

Let us provide some concrete examples of gapless Hamiltonians with the above symmetries.
An example of a gapless Hamiltonian with $\U(1)\times\U(1)$ and $\Rep(D_8)$ symmetries is given by
\begin{equation}
  H_{\U(1)_4^2}=\sum_{j}\left(Y_{j}Z_{j+2}-Z_{j}Y_{j+2}\right).
\end{equation}
We note that this Hamiltonian is equivalent to two copies of the Hamiltonian \eqref{eq:U(1)_4} defined on odd and even sites separately.
A gapless Hamiltonian with $\U(1)\times\U(1)$ and $\mathbb{Z}_2^3$ symmetry with the type III anomaly is obtained by applying the KT transformation to the above Hamiltonian. The resulting Hamiltonian flows to the $\mathrm{Spin}(4)_1$ WZW CFT in the low-energy limit and is given by
\begin{equation}
  H_{\mathrm{Spin}(4)_1}=\sum_{j}\left(Y_{j}X_{j+1}Z_{j+2}-Z_{j}X_{j+1}Y_{j+2}\right).
\end{equation}
To see why this Hamiltonian flows to the $\mathrm{Spin}(4)_1$ WZW CFT, we use the Jordan--Wigner transformation with respect to the diagonal $\mathbb{Z}_2$ symmetry to rewrite the Hamiltonian in terms of Majorana operators.
We then obtain the following free Majorana Hamiltonian:
\begin{equation}
  H_{\mathrm{Spin}(4)_1}^{F}=\sum_{j}\left(ia_{j}a_{j+2}+ib_{j}b_{j+2}\right).
\end{equation}

\section{Outlook}\label{sec:outlook}
In this paper, we introduced the notion of a symmetry span and showed that incompatible constraints from multiple symmetry embeddings can enforce gaplessness, even when the individual symmetries are non-anomalous in the UV.
We provided concrete examples in 1+1 dimensions, and constructed lattice realizations that flow to gapless theories with emergent anomalous continuous symmetries in the IR.

In all continuum examples studied, the full symmetry $\Sfaith$ contains an anomalous continuous symmetry whose anomaly by itself enforces gaplessness.
It would be interesting to determine whether this is always the case for theories satisfying the span condition, or whether the span can enforce gaplessness in a genuinely new way---without any anomalous continuous symmetry present.

On the lattice side, several open problems remain.
The lattice models presented in this paper realize the symmetry span only at special points (e.g.\ the free-fermion point of the torus CFT); finding lattice realizations at other points on the moduli space, particularly for genuinely interacting models, is an important challenge.
It would also be interesting to characterize the full symmetry algebra generated by the span operators---beyond the Onsager subalgebra identified in Sec.~\ref{sec:lattice_realization}---and to understand how the emergent anomalous continuous symmetry can be diagnosed directly at the lattice level.

Meanwhile, we have some established mechanisms of dynamical consequences inherent to lattice systems, such as the Lieb--Schultz--Mattis (LSM) type theorem and the SPT-LSM type argument \cite{Lu:2017ego, Yang:2017frp, Lu:2017mmn,Else:2019lft, Jiang:2019ryo, Pace:2024acq}.
Studying gaplessness arguments based on symmetry spans in relation to these mechanisms is an interesting future direction.

Finally, the gaplessness argument based on symmetry spans is not restricted to 1+1 dimensions: the pullback condition \eqref{eq:gapped condition} applies whenever symmetries are described by appropriate higher categories.
Constructing explicit symmetry spans, theories, and lattice realizations in $d>2$ spacetime dimensions is a natural next step.

\paragraph*{Acknowledgments.}
T.A.~is supported by JST CREST (Grant No.~JPMJCR19T2) and JSPS KAKENHI Grant Number 25KJ1557.
K.O.~is supported by JSPS KAKENHI Grant-in-Aid No.22K13969 and No.24K00522.
K.O.~also acknowledges support by the Simons Foundation Grant \#888984 (Simons Collaboration on Global Categorical Symmetries).

\appendix
\section{Bosonization and fermionization}\label{sec:bosonization}
Here we briefly summarize bosonization and fermionization in two spacetime dimensions \cite{Karch:2019lnn, Ji:2019jhk}.
For a bosonic theory with a non-anomalous $\mathbb{Z}_2$ zero-form symmetry, we can define its fermionization, which maps the bosonic theory to a fermionic theory.\footnote{In this paper, a bosonic theory is defined on oriented spacetime manifolds, while a fermionic theory is defined on spin manifolds.}
One can also define bosonization as the inverse operation of fermionization.
Let us denote bosonic and fermionic theories by $\mathsf{D}$ and $\mathsf{F}$, respectively.
We first define our bosonization and fermionization as follows:
\begin{align}
    Z_{\mathsf{Fer}(\mathsf{D})}[A+\rho]&\coloneqq \#\sum_{a}Z_{\mathsf{D}}[a](-1)^{\arf(a+\rho)+\arf(\rho)+\int aA},\\
    Z_{\mathsf{Bos}(\mathsf{F})}[A]&\coloneqq \#\sum_{a}Z_{\mathsf{F}}[a+\rho](-1)^{\arf(A+\rho)+\arf(\rho)+\int aA}.
\end{align}
Here, $\rho$ indicates the choice of the spin structure, i.e., its derivative is equal to the second Stiefel--Whitney class $\delta\rho=w_2$.
The quantity $\arf(\rho)$ is the Arf invariant of two-dimensional spacetime manifolds, and $\#$ stands for numerical factors that depend on the topology of the spacetime manifold.
We also use slightly different operations, which we call the Jordan--Wigner transformation and its inverse:
\begin{align}
    Z_{\mathsf{JW}(\mathsf{D})}[A+\rho]&\coloneqq \#\sum_{a}Z_{\mathsf{D}}[a](-1)^{\arf(A+a+\rho)},\\
    Z_{\mathsf{JW}^{-1}(\mathsf{F})}[A]&\coloneqq \#\sum_{a}Z_{\mathsf{F}}[a+\rho](-1)^{\arf(A+a+\rho)}.
\end{align}
The following relations are useful for forward calculations:
\begin{gather}
  \begin{split}
    \arf(a+A+\rho)=\arf(a+\rho)+\arf(A+\rho)\\
    +\arf(\rho)+\int aA \mod 2,
  \end{split}\\
  \#\sum_{a}(-1)^{\arf(a+\rho)+\arf(\rho)+\int aA}=(-1)^{\arf(A+\rho)}.
\end{gather}
Consider a bosonic theory $\mathsf{D}$ with a non-anomalous $\mathbb{Z}_2\times\mathbb{Z}_2$ symmetry coupled to background fields $A,B$.
The partition function of the Kennedy--Tasaki transformed theory is given by
\begin{equation}
  Z_{\mathsf{KT}(\mathsf{D})}[A,B]=\#\sum_{a,b}Z_{\mathsf{D}}[a,b](-1)^{\int ab+aB+bA+AB}.
\end{equation}
One can show that this transformation is equivalent to the following sequence of Jordan--Wigner transformations:
\begin{equation}\label{eq:KT_JW}
  \mathsf{KT}=\mathsf{JW}^{-1}\circ\mathsf{JW}_{1,2}.
\end{equation}
Here, $\mathsf{JW}_{1,2}$ is the Jordan--Wigner transformation with respect to the first and second $\mathbb{Z}_2$ symmetries, and $\mathsf{JW}^{-1}$ is the inverse Jordan--Wigner transformation with respect to the diagonal fermion parity symmetry.
The equivalence can be shown by straightforward calculations as follows:
\begin{widetext}
  \begin{align}
  \begin{split}
    Z_{\mathsf{JW}^{-1}\circ\mathsf{JW}_{1,2}(\mathsf{D})}[A^\prime,B^\prime]
    &=\#\sum_{a,b,\tilde{a}}Z_{\mathsf{D}}[a,b](-1)^{\arf(A+\tilde{a}+\rho)}
    (-1)^{\int \arf(\tilde{a}+a+\rho)+\arf(\tilde{a}+B^\prime+b+\rho)}\\
    &=\#\sum_{a,b,\tilde{a}}Z_{\mathsf{D}}[a,b](-1)^{\arf(A^\prime+\rho)+\arf(\tilde{a}+\rho)+\arf(\rho)+\int\tilde{a}A^\prime}
    (-1)^{\arf(a+\rho)+\arf(B+b+\rho)+\int \tilde{a}(a+B^\prime+b)}\\
    &=\#\sum_{a,b}Z_{\mathsf{D}}[a,b](-1)^{\arf(A^\prime+\rho)+\arf(a+\rho)+\arf(B^\prime+b+\rho)+\arf(a+b+A^\prime+B^\prime+\rho)}\\
    &=\#\sum_{a,b}Z_{\mathsf{D}}[a,b](-1)^{\int bB^\prime+a(b+A^\prime+B^\prime)+b(A^\prime+B^\prime)+A^\prime B^\prime}\\
    &=\#\sum_{a,b}Z_{\mathsf{D}}[a,b](-1)^{\int ab+a(A^\prime+B^\prime)+bA^\prime+A^\prime B^\prime}\\
  \end{split}
\end{align}
\end{widetext}
By redefining the quantum $\mathbb{Z}_2^2$ symmetry as $A=A^\prime$ and $B=A^\prime+B^\prime$, we obtain the desired equivalence. Note that $(-1)^{\int A^2}=1$ for any $\mathbb{Z}_2$ gauge field $A$ on oriented manifolds.

\section{Type III anomaly in \texorpdfstring{$\mathrm{Spin}(4)_1$}{Spin(4)1} WZW CFT}\label{sec:spin(4)}
\subsection{Symmetry on lattice}
\begin{figure}[tbhp]
  \centering
\begin{tikzcd}[row sep=4em, column sep=5em]
\U(1)_{4}^{2}
  \arrow[r, "{\mathsf{JW}_{1,2}}"']
  \arrow[rr, bend left=25, "{\mathsf{KT}}"]
&
\mathrm{Dirac}^{2}
  \arrow[r, "{\mathsf{JW}^{-1}}"']
  \arrow[d, <->, "{\times\,\mathsf{Arf}}"]
&
\mathrm{Spin}(4)_{1}
\\
&
\widetilde{\mathrm{Dirac}^{2}}
  \arrow[ru, "{\mathsf{Bos}}"']
  \arrow[r, "{\mathsf{JW}^{-1}}"']
&
\widetilde{\mathrm{Spin}(4)_{1}}
  \arrow[u, <->, "{\text{gauging}}"']
\end{tikzcd}
\caption{Relations of several CFTs and gauging operations.}\label{fig:Spin(4)_web}
\end{figure}
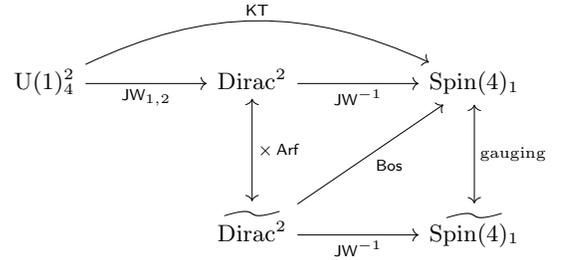

To identify how the symmetry span $\mathbf{Vect}_{\U(1)^2}\hookleftarrow\Vect_{\mathbb{Z}_2^2}\hookrightarrow\Vect_{\mathbb{Z}_2^3}^\omega$ is realized in the $\mathrm{Spin}(4)_1$ WZW CFT, we use its lattice realization.
We first consider two copies of the Dirac fermion chain defined by the Hamiltonian
\begin{equation}
  H_{\mathrm{Dirac}^2}=i\sum_{j}\left(a_{j}a_{j+2}+b_{j}b_{j+2}\right).
\end{equation}
The two $\U(1)$ charges of this system are given by
\begin{equation}
  Q_{\mathrm{Dirac}^2}^{(1)}=\frac{1}{2}\sum_{j:\text{odd}}ia_{j}b_{j},\quad 
  Q_{\mathrm{Dirac}^2}^{(2)}=\frac{1}{2}\sum_{j:\text{even}}ia_{j}b_{j}.
\end{equation}
These $\U(1)$ charges come from the following $\U(1)$ charges in the $\U(1)_4^2$ and the $\mathrm{Spin}(4)_1$ WZW CFT:
\begin{equation}\label{eq:Dirac_U(1)_charge}
  Q_{\U(1)_4^2}^{(1)}=\frac{1}{2}\sum_{j:\text{odd}}X_{j},\quad
  Q_{\U(1)_4^2}^{(2)}=\frac{1}{2}\sum_{j:\text{even}}X_{j}.
\end{equation}
Here, $\U(1)_4$, $\mathrm{Spin}(4)_1$, and $\mathrm{Dirac}^2$ are related by Jordan--Wigner transformation maps, as in Fig.~\ref{fig:Spin(4)_web}.

On the other hand, the two $\U(1)$ charges of the fermionized theory of the $\mathrm{Spin}(4)_1$ WZW CFT, denoted by $\widetilde{\mathrm{Dirac}^2}$, are given by
\begin{equation}\label{eq:Dirac_U(1)_charge_2}
  Q_{\widetilde{\mathrm{Dirac}^2}}^{(1)}=\frac{1}{2}\sum_{j:\text{odd}}ia_{j}a_{j+1},\quad
  Q_{\widetilde{\mathrm{Dirac}^2}}^{(2)}=\frac{1}{2}\sum_{j:\text{even}}-ib_{j}b_{j+1}.
\end{equation}
For the derivation of these $\U(1)$ charges, we use the following JW transformation \cite{Pace:2024oys}\footnote{In \cite{Pace:2024oys}, the authors call it fermionization, but we are using a different terminology.}:
\begin{equation}
  X_{j}\mapsto ia_{j}b_{j},\quad Z_{j}Z_{j+1}\mapsto 
  \begin{cases*}
    -ia_{j}a_{j+1}~~j:\text{odd},\\
    -ib_{j}b_{j+1}~~j:\text{even}.
  \end{cases*}
\end{equation}
Let us define another $\U(1)$ charge as follows:
\begin{align}\label{eq:Dirac_U(1)_charge_3}
  \begin{split}
    \mathcal{Q}_{\widetilde{\mathrm{Dirac}^2}}^{(2)}
    &\coloneqq T_{b}^{\text{(even)}}Q_{\widetilde{\mathrm{Dirac}^2}}^{(2)}(T_{b}^{\text{(even)}})^{-1}\\
    &\coloneqq \frac{1}{2}\sum_{j:\text{even}}-ib_{j+1}b_{j}=\frac{1}{2}\sum_{j:\text{odd}}ib_{j}b_{j+1},
  \end{split}
\end{align}
where $T_{b}^{\text{(even)}}$ is the translation operator for Majorana operators on even sites.

The $\mathbb{Z}_2$ symmetry with which the two $\mathbb{Z}_2$ symmetries carry the type III anomaly in the $\mathrm{Spin}(4)_1$ WZW CFT is generated by
\begin{equation}
  U_{c}=\prod_{j}\exp\left(\frac{\pi i}{4}(-1)^jZ_{j}Z_{j+1}\right),
\end{equation}
and it is mapped to the following $\mathbb{Z}_2$ symmetry in the $\widetilde{\mathrm{Dirac}^2}$ theory:
\begin{equation}\label{eq:Dirac_Z2_sym}
  U_{c}^f=\prod_{j}\exp\left(\frac{\pi i}{4}(-ia_{2j}b_{2j}+ia_{2j-1}b_{2j-1})\right).
\end{equation}
Upon fermionization, the type III anomaly is mapped to a $D_8^f$ symmetry, where the central $\mathbb{Z}_2$ is the fermion parity symmetry.

\subsection{Symmetry under bosonization}
In the previous subsection, we specified the symmetry operators for the span $\U(1)^2\hookleftarrow \mathbb{Z}_2^2\hookrightarrow D_8^f$ in the fermion theory $\widetilde{\mathrm{Dirac}^2}$.
To proceed, we use the isomorphism between $\SO(4)$ and $\SU(2)\times\SU(2)/\mathbb{Z}_2$.
Specifically, we identify elements of $\SU(2)$ with unit quaternions $\mathbb{H}$ and represent elements of $\SO(4)$ as pairs of unit quaternions $(u,v)\in\SU(2)\times\SU(2)$ that act on $\mathbb{H}$ as
\begin{equation}
  w\mapsto uwv,\quad w\in\mathbb{H}.
\end{equation}
Since $\mathbb{H}\cong\mathbb{R}^4$, this action defines an element of $\SO(4)$.
More specifically, when we write the infinitesimal generators of $\SU(2)^2$ as
\begin{gather}\label{eq:SU2_infinitesimal}
    \begin{split}
      u=  1
          + \frac{dt_{ab} + dt_{cd}}{2}\, i
          + \frac{dt_{ac} - dt_{bd}}{2}\, j
          + \frac{dt_{ad} + dt_{bc}}{2}\, k,\\
      v=  1
          + \frac{dt_{ab} - dt_{cd}}{2}\, i
          + \frac{dt_{ac} + dt_{bd}}{2}\, j
          + \frac{dt_{ad} - dt_{bc}}{2}\, k,
    \end{split}
\end{gather}
were $i,j,k$ are the imaginary units of quaternions satisfying $i^2=j^2=k^2=ijk=-1$, and   
 the corresponding $\SO(4)$ matrix is given by
\begin{equation}\label{eq:SO(4)_infinitesimal}
    u \,w\, v
=
\begin{pmatrix}
1        & -dt_{ab} & -dt_{ac} & -dt_{ad} \\
dt_{ab}  & 1        & -dt_{bc} & -dt_{bd} \\
dt_{ac}  & dt_{bc}  & 1        & -dt_{cd} \\
dt_{ad}  & dt_{bd}  & dt_{cd}  & 1
\end{pmatrix}
\begin{pmatrix}
a \\ b \\ c \\ d
\end{pmatrix}.
\end{equation}

In the $\widetilde{\mathrm{Dirac}^2}$ theory, we have an $\SO(4)$ symmetry generated by rotations of four Majorana fermions.
Let us see how to regard the $\U(1)^2$ symmetry \eqref{eq:Dirac_U(1)_charge_2} and the $\mathbb{Z}_2$ symmetry \eqref{eq:Dirac_Z2_sym} as elements of $\SU(2)\times\SU(2)/\mathbb{Z}_2$.
First, consider the $\U(1)$ charge $Q_{\widetilde{\mathrm{Dirac}^2}}^{(1)}$ and another $\U(1)$ charge $\mathcal{Q}_{\widetilde{\mathrm{Dirac}^2}}^{(2)}$ in \eqref{eq:Dirac_U(1)_charge_3}.
We write the four Majorana fermions as $(a_{2j-1},b_{2j-1},a_{2j},b_{2j})^{\mathsf{T}}$.
Then the $\U(1)$ symmetries generated by $Q_{\widetilde{\mathrm{Dirac}^2}}^{(1)}$ and $\mathcal{Q}_{\widetilde{\mathrm{Dirac}^2}}^{(2)}$ act on the Majorana fermions as
\begin{gather}
    \begin{pmatrix}
        a_{2i-1} \\ b_{2i-1} \\ a_{2i} \\ b_{2i}
    \end{pmatrix}
    \xrightarrow{Q_{\widetilde{\mathrm{Dirac}^2}}^{(1)}}
    \begin{pmatrix}
        \cos\theta & 0 & \sin\theta & 0 \\
        0          & 1 & 0 & 0 \\
        -\sin\theta & 0 & \cos\theta & 0 \\
        0          & 0 & 0 & 1
    \end{pmatrix}
    \begin{pmatrix}
        a_{2i-1} \\ b_{2i-1} \\ a_{2i} \\ b_{2i}
    \end{pmatrix},\\
    \begin{pmatrix}
        a_{2i-1} \\ b_{2i-1} \\ a_{2i} \\ b_{2i}
    \end{pmatrix}
    \xrightarrow{\mathcal{Q}_{\widetilde{\mathrm{Dirac}^2}}^{(2)}}
    \begin{pmatrix}
        1 & 0 & 0 & 0 \\
        0 & \cos\theta & 0 & \sin\theta \\
        0 & 0 & 1 & 0 \\
        0 & -\sin\theta & 0 & \cos\theta
    \end{pmatrix}
    \begin{pmatrix}
        a_{2i-1} \\ b_{2i-1} \\ a_{2i} \\ b_{2i}
    \end{pmatrix}.
\end{gather}
Comparing these actions with the infinitesimal $\SO(4)$ rotation \eqref{eq:SO(4)_infinitesimal}, we find that these $\U(1)$ symmetries correspond to the following elements of $\SU(2)\times\SU(2)/\mathbb{Z}_2$:
\begin{equation}
    (u,v)=
    \begin{cases}
        (-\frac{d\theta}{2}j,-\frac{d\theta}{2}j) & \text{for } Q_{\widetilde{\mathrm{Dirac}^2}}^{(1)},\\
        (+\frac{d\theta}{2}j,-\frac{d\theta}{2}j) & \text{for } \mathcal{Q}_{\widetilde{\mathrm{Dirac}^2}}^{(2)}.
    \end{cases}
\end{equation}
Under the bosonization map, these $\U(1)$ symmetries are mapped to the diagonal and anti-diagonal shift symmetries in the $\mathrm{Spin}(4)_1\cong \mathrm{SU}(2)_1\times \mathrm{SU}(2)_1$ WZW CFT, respectively.
Recall that the $\U(1)$ charge $Q_{\widetilde{\mathrm{Dirac}^2}}^{(2)}$ in \eqref{eq:Dirac_U(1)_charge_2} is related to $\mathcal{Q}_{\widetilde{\mathrm{Dirac}^2}}^{(2)}$ by the translation of even-site Majorana fermions, as in \eqref{eq:Dirac_U(1)_charge_3}.
Since the translation operator for even-site Majorana fermions flows to the $\mathbb{Z}_2$ chiral charge conjugation for one of the Dirac fermions in the low-energy limit, we find that the $\U(1)$ symmetry generated by $Q_{\widetilde{\mathrm{Dirac}^2}}^{(2)}$ is mapped to the anti-diagonal $\U(1)$ winding symmetry in the $\SU(2)_1\times\SU(2)_1$ WZW CFT.
To see this, we identify each chiral Dirac fermion as
\begin{align}
  \begin{split}
    \psi_{L/R}^{(1)}&\sim a_{4j-1}\pm a_{4j+1},\\
    \psi_{L/R}^{(2)}&\sim b_{4j-1}\pm b_{4j+1},\\
    \psi_{L/R}^{(3)}&\sim a_{4j}\pm a_{4j+2},\\
    \psi_{L/R}^{(4)}&\sim b_{4j}\pm b_{4j+2}.
  \end{split}
\end{align}
Then the translation for even-site Majoranas acts on $\psi_{R}^{(3)}\mapsto -\psi_{R}^{(3)}$, and so the precise correspondence of $(u,v)$ for $Q_{\widetilde{\mathrm{Dirac}^2}}^{(2)}$ is given by
\begin{gather}
    \begin{split}
      (\frac{d\theta}{2} j,-\frac{d\theta}{2} j)\quad \text{(for left mover)},\\
      (-\frac{d\theta}{2} j,\frac{d\theta}{2} j)\quad\text{(for right mover)},
    \end{split}
\end{gather}
which corresponds to the anti-diagonal winding symmetry in the $\SU(2)_1\times\SU(2)_1$ WZW CFT.

Finally, looking at the symmetry operator $U_c^f$ in \eqref{eq:Dirac_Z2_sym}, we find that it corresponds to $(1,-j)\in \SU(2)\times\SU(2)/\mathbb{Z}_2$, and it is mapped to the $\mathbb{Z}_2$ symmetry that acts as charge conjugation for one of the $\SU(2)_1$ WZW CFTs in the $\mathrm{Spin}(4)_1\cong \SU(2)_1\times\SU(2)_1$ WZW CFT.

To summarize, for the span $\mathbf{Vect}_{\U(1)^2}\hookleftarrow\Vect_{\mathbb{Z}_2^2}\hookrightarrow\Vect_{\mathbb{Z}_2^3}^\omega$ in the $\mathrm{Spin}(4)_1$ WZW CFT, the $\U(1)^2$ symmetry corresponds to the diagonal shift and anti-diagonal winding symmetries in the $\mathrm{Spin}(4)_1=\SU(2)_1\times\SU(2)_1$ CFT, and the other $\mathbb{Z}_2$ symmetry in the type III anomaly corresponds to charge conjugation for one of the $\SU(2)_1$ CFTs.
These identifications are consistent with the continuum example in Sec.~\ref{subsec:T2_CFT}.
We note that the last $\mathbb{Z}_2$ symmetry does not commute with the $\U(1)^2$ symmetry, and this is consistent with both continuum and lattice analyses.

\bibliography{ref}

\end{document}